\documentclass[prb,twocolumn,showpacs,amsmath]{revtex4}
\usepackage{amssymb}
\usepackage{amsmath}
\usepackage{graphicx}
\usepackage{color}

\def\be{\begin{equation}}
\def\ee{\end{equation}}

\newcommand{\combinatorial}[2]{
\begin{pmatrix}
#1\\
#2
\end{pmatrix}
}
\usepackage{hyperref}

\begin{document}
\title{Numerical cookbook for electronic quantum transport at finite frequency}
\date{\today}
\author{Oleksii Shevtsov}
\email{oleksii.shevtsov@cea.fr}
\author{Xavier Waintal}
\email{xavier.waintal@cea.fr}
\affiliation{Commissariat a l'Energie Atomique, DSM/INAC/SPSMS/GT, 17 rue des Martyrs, 38054
Grenoble Cedex 9, France}
\begin{abstract}
Building on the many existing algorithms for calculating the DC transport properties of quantum tight-binding models, we develop a systematic approach that expresses finite frequency observables in terms of the stationary Green's function of the system, i.e. the natural output of most DC numerical codes. Our framework allows to extend the simulations capabilities of existing codes to a large class of observables including, for instance, AC conductance, quantum capacitance, quantum pumping, spin pumping or photo-assisted shot noise.
The theory is developed within the framework of Keldysh formalism and we provide explicit links with the alternative (and equivalent) scattering approach. We illustrate the formalism with a study of the AC conductance in a quantum point contact and an electronic Mach-Zehnder interferometer in the quantum Hall regime.
\end{abstract}

\maketitle

\section{Introduction}
Numerical simulations of the transport properties of quantum tight-biding models are now a mature field and are routinely used either alone or in conjunction with ab-initio simulations and/or electrostatic simulations. Examples include a wide variety of systems ranging from graphene devices\cite{Wallace_graphene,Reich_Thomsen_TB_graphene,Novoselov_review,DasSarma_transport_graphene_review,CastroNeto_TB_uniaxial_strain,Khorasani_TB_patterned_graphene}, quantum Hall effect \cite{Hofstadter,Chalker_models_of_QHE,Gong_TB_QHE,Koshino_3D_TB_QHE} or spintronic devices \cite{Aharony_spintronic,Cohen_spintronic,Virgilio_spintronic} to topological insulators \cite{Kane_Mele2005,Kane_Mele2005_Z2,Fu_Kane_Mele,Mao_TB_TI}, semi-conducting nanowires \cite{Niquet_nanowires,Bescond_Ge_nanowires,Niquet_TB_vs_KP} or hybrid superconducting systems \cite{Kawabata_hybrid,Lambert_hybrid,Arrachea_hybrid,Xiong_hybrid}.

These simulations are based on two equivalent approaches, the scattering (or Landauer-B\"uttiker) formalism \cite{Buttiker1988,Pastawski1991,Imry_Landauer,Landauer,Datta1997} and the Non Equilibrium Green's Function formalism \cite{Keldysh_original,Rammer_Smith,Rammer2007,Jauho2010} (NEGF, itself based on the Keldysh formalism). The development of the numerical methods for simulating quantum transport is perhaps as old as those formalisms \cite{Sullivan_numerical_transport_old,William_numerical_transport_old,Bandy_numerical_transport_old,Jensen_numerical_transport_old}.  One of the most popular algorithms is the Recursive Green's Function method \cite{Rotter_recursive_GF,Drouvelis_recursive_GF,Kahen_recursive_GF} and its generalizations for tackling larger systems, complex geometries and multiterminal measurements \cite{Waintal_KNIT}. Typical observables accessible with those techniques are the conductance, the noise, as well as local properties like electronic density, current density or spin currents.

In recent years, it has become clear that probing quantum systems at finite frequencies can provide entirely new physical insights, and finite frequency mesoscopic physics is emerging as a field on its own. Among the pioneer works, the extension of the scattering approach to finite frequency initiated in \cite{Pastawski1992,Buttiker_JOP1993,Buttiker_PRL1993,Buttiker_PLA1993,Buttiker_PRL1993_1,Buttiker_Nuovo_Cimento,Anantram_Datta_PhaseBreaking_on_AC_Mesoscopic} has led to many interesting predictions including the quantum capacitance, finite frequency shot noise, quantum pumping, spin pumping, etc. On the experimental level, apart from the huge corpus of work with quantum-bits (semi-conductor or superconducting based), one observes an increasing interest in high frequency physics, including several striking experiments on quantum capacitance \cite{Xu_quantum_capacitance,Droscher_quantum_capacitance,Lorke_quantum_capacitance}, quantum inductance \cite{Asakawa_quantum_inductance,Begliarbekov_quantum_inductance}, or electronic quantum optics \cite{Ji2003Nat,Heiblum_PRL_MZ,Heiblum_PRL07,Heiblum_NatPhys07,Roulleau_PRB07,Roulleau_PRL08,Roulleau_PRL08bis,Roulleau_PRL09,Oliver_Science_HBT,Camino_AB_in_QHE,Kotimaki_AB_rings}.

On the numerical side, finite frequency physics has attracted comparatively limited interest so far. Pioneer works include some calculations by Guo \textit{et al} \cite{Guo_Current_partition,Guo_AC_conductance_NTs,Guo_dyn_cond_mes_waveguides,Guo_NonAdiab_charge_pump_Multiphoton_expansion} as well as a few others \cite{Tien_Gordon,Jauho_Wingreen_Meir,Wingreen_Jauho_Meir,Anantram_Datta_PhaseBreaking_on_AC_Mesoscopic,Zheng_LB_formula_TD_trans_NEGF,Aguado_Platero,Cai_SelfCons_AC_NEGF,Petitjean_Semiclass_AC_Chaotic,Wei_Wang_Curr_cons_NEGF_AC_Trans,Kienle_SelfCons_AC_in_device_NEGF}.

The aim of this paper is to provide a comprehensive approach to numerical simulations of quantum transport at finite frequency, bridging the gap between the output of the existing numerical codes (typically the Retarded Green's function of a sub part of the system) and the actual AC observables. We shall provide a systematic way to derive formulas that express those AC quantities (say the AC conductance) as the trace of a product of a few easily accessible Green's functions. Some of these formulas are already known in the literature, others are new. More specifically, in presence of a perturbation of the form
$V_{ac} \cos \omega t$ we provide two complementary approaches. The first one is a diagrammatic technique for a systematic expansion of the retarded Green's function in powers of $V_{ac}$. It allows to easily recover the (mostly known) lowest order expressions but also to go beyond linear response or the wide band limit using a simple set of Feynman like rules. The second method allows to access the opposite adiabatic limit ($\omega\rightarrow 0$) and systematically expand any observable in powers of $\omega$. 

The paper is organized as follows. Section \ref{DC_primer} contains an introduction to numerical simulations of stationary (DC) quantum transport. We focus in particular on the output of the DC codes that will be the input for the calculations of the finite frequency (AC) quantities. We are then ready, in Section \ref{Cookbook}, to present a rather long list of AC expressions that relate one AC quantity (such as AC conductance or photocurrent) to the integral (over energy) of the trace of a product of DC Green's functions. For simplicity we give these expressions in the so called wide band limit and defer the full expressions (that should be used in the numerics) to the Appendix. All the expressions given in the Appendix (but the first one) are new to the best of our knowledge. In the beginning of Section \ref{Ch_Applications} we provide some technical details explaining how to evaluate the energy integrals in practice. Further in this section we consider two applications of increasing complexity. In the first one, we consider the AC conductance in a point contact geometry and show that such a measurement would allow to extract the time of flight of the electrons through the device. In the second, we consider the  full scale simulation of an electronic Mach-Zehnder interferometer made out of the edge states of a two dimensional gas  in the quantum Hall regime.  

The full machinery used to derive the expressions of Section \ref{Cookbook} (and the Appendix) is developed in Sections \ref{Ch_TD_NEGF}-\ref{Ch_Pert_leads}. It provides a set of simple rules allowing to derive the expressions given in Section \ref{Cookbook} as well as any other similar expression beyond those given explicitly here. Section \ref{Ch_TD_NEGF} briefly introduces the necessary notations and main results of the Keldysh formalism. Section \ref{Ch_Periodic_potential} focuses on the particular case where the perturbation is periodic in time. We then proceed with developing systematic perturbative expansions around two distinct limits: first we provide a diagrammatic technique to expand the results in powers of the amplitude of the perturbation. Second, we expand around the adiabatic limit where the frequency of the AC perturbation is very small. These expansions are presented in Section \ref{Ch_Pert_device} (perturbation applied inside the system) and Section\ref{Ch_Pert_leads} (perturbation in the electrodes). 

\section{Building block for AC transport: The Green's functions of the stationary problem}\label{DC_primer}

\begin{figure}[h]
\includegraphics[scale=0.50]{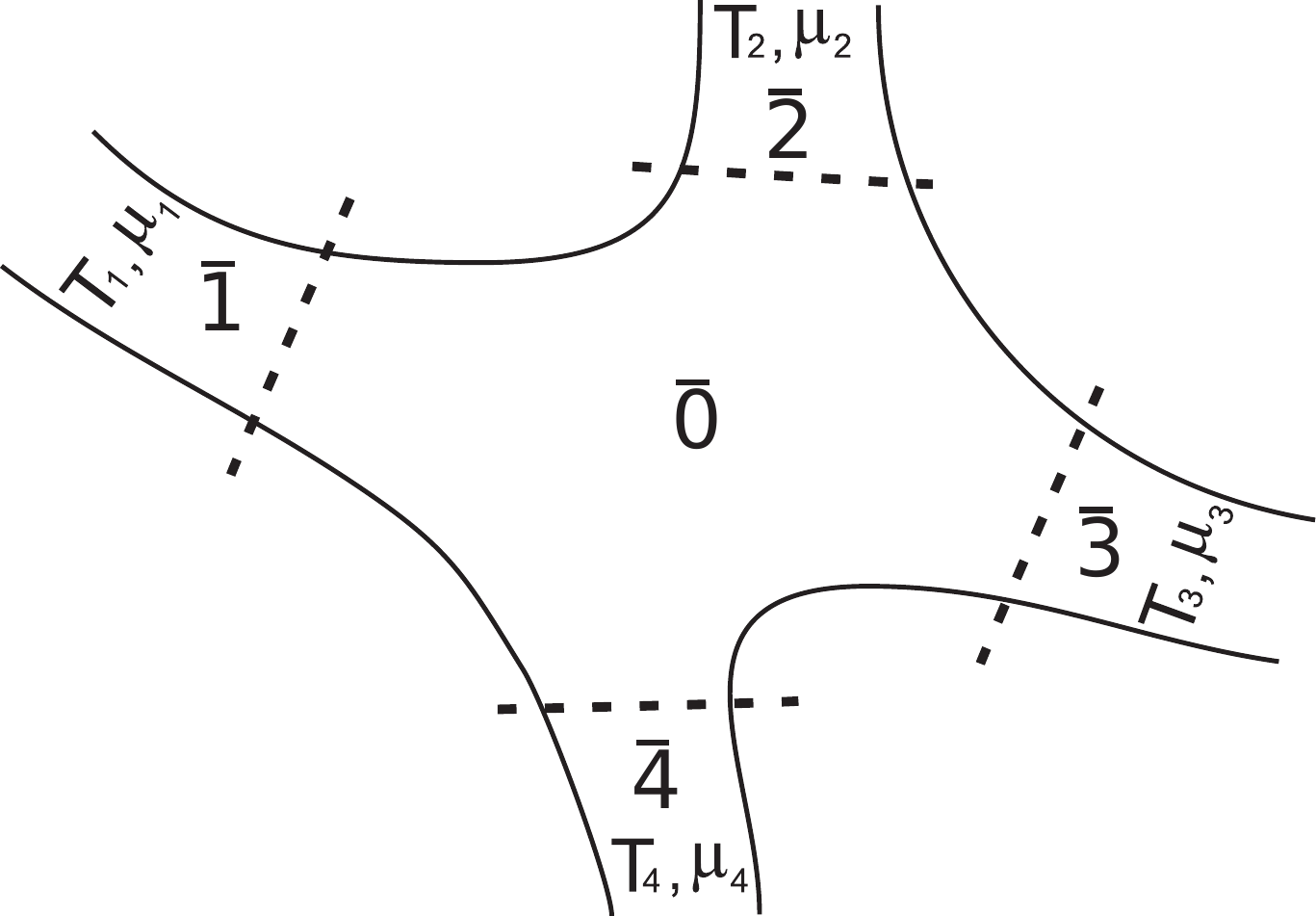}
\caption{Sketch of a multi-terminal mesocopic system connected to $M=4$ leads. Numbers with a bar on top denote the parts of our system, i.e. $\bar{0}$ for the device region and $\bar{1}...\bar{4}$ for the leads.}\label{System}
\end{figure}

We begin this manuscript with a brief account of the numerical methods used for simulating the stationary (DC) transport properties of quantum systems. We focus on tight-binding models within the NEGF formalism\cite{Wingreen_Meir} which has become, perhaps, the standard approach for this problem thanks to the development of recursive \cite{Rotter_recursive_GF,Drouvelis_recursive_GF,Kahen_recursive_GF} and other advanced algorithms \cite{Waintal_KNIT,Mamaluy2003,Mamaluy2005}. The basic objects obtained as the {\it output} of those numerical tools are the retarded  Green's functions of the system which will be the {\it input } of the expressions for the AC observables derived later in this manuscript.

We consider a generic quadratic discrete Hamiltonian of an open system written as
\be
\label{dc_Hamiltonian}
{\hat{\mathbf{H}}}=
\sum_{n,m}{\mathbf{H}}_{nm} c^{\dagger}_nc_m.
\ee
where $c^{\dagger}_n$ ($c_n$) are the usual creation (destruction) operator on site $n$. The site index $n$ is very generic and includes all the degrees of freedom present in the system (spatial, momentum, spin, electron/hole, orbital). In terms of the (infinite) $\mathbf{H}$ matrix, the retarded Green's function is defined as,
\begin{align}
\label{G_0_definition1}
\mathcal{G}(E)=\left(E+i\epsilon-\mathbf{H} \right)^{-1},
\end{align}
where $\epsilon$ is an infinitely small positive number. We now focus on the particular geometry of our mesoscopic system: a finite region, referred as  $\bar{0}$ connected to $M$ semi-infinite electrodes $\bar{1}...\bar{M}$, see Fig.~\ref{System}. In the rest of this manuscript, notations such as $A_{\bar{m}\bar{m}^{\prime}}$ always refer to the corresponding sub-block of the full infinite matrix $A$. Each electrode is kept at equilibrium with chemical potential $\mu_m$ and temperature $T_m$. The first step in the NEGF formalism consists in integrating
out the electrodes degrees of freedom. One obtains for the retarded Green function inside the device region 
$\mathcal{G}_{\bar{0}\bar{0}}(E)$:
\begin{align}
\label{G_0_definition}
\mathcal{G}_{\bar{0}\bar{0}}(E)=\left(E-H-\sum_{m=1}^M\mathsf{\Sigma}^r(m;E)\right)^{-1},
\end{align}
where $H=\mathbf{H}_{\bar{0}\bar{0}}$ is the Hamiltonian matrix projected inside the device region (see Fig.~\ref{System})
and $\mathsf{\Sigma}^r(m;E)$ is the (retarded) self-energy due to the presence of the lead $m$. The latter is given by,
\be
\label{SigmaR}
\mathsf{\Sigma}^r(m;E) = \mathbf{H}_{\bar{0}\bar{m}} (E +i\epsilon -\mathbf{H}_{\bar{m}\bar{m}})^{-1} \mathbf{H}_{\bar{m}\bar{0}}.
\ee

Equations (\ref{G_0_definition}) and (\ref{SigmaR}) are the typical raw output of, say, recursive techniques\cite{Rotter_recursive_GF,Drouvelis_recursive_GF,Kahen_recursive_GF,Waintal_KNIT,Mamaluy2003,Mamaluy2005} from which one can calculate various physical observables such as conductance or current noise. For instance, the celebrated Landauer formula for the current flowing from lead $m$ reads in this context\cite{Wingreen_Meir,Datta1997},
\begin{align}
\label{landauer}
I_m=\frac{e}{h}\int dE \sum_{m'=1}^{M}\left(f_m-f_{m'}\right)\mathrm{Tr}\left[\mathcal{G}_0\Gamma_{m'}\mathcal{G}_0^{\dagger}\Gamma_{m}\right],
\end{align}
where we have introduced the Fermi function $f_m(E)=1/(e^{(E-\mu_m)/kT_m}+1)$, the standard broadening matrix
\begin{align}
\label{Gamma_def}
\Gamma_m(E)=i\left[\mathsf{\Sigma}^{r}(m;E)-\mathsf{\Sigma}^{r\dagger}(m;E)\right].
\end{align}
and the shortcut
\begin{equation}
\mathcal{G}_l(E)\equiv\mathcal{G}_{\bar{0}\bar{0}}(E+\frac{\hbar\omega l}{2}).
\label{Eq_GF_notation}
\end{equation}

In what follows, our aim is to extend expressions such as Eq.(\ref{landauer}) to AC quantities. Those expressions will involve photon absorption and emission processes and therefore will require the calculation of $\mathcal{G}_l(E)$ for $l\ne 0$. 

\section{Cookbook for calculating AC observables}\label{Cookbook}

This section is intended as some sort of dictionary where we provide expressions for various finite frequency response functions. Those expressions allow to extend directly DC numerical codes to the AC regime. They are given without derivation and we refer to Sections \ref{Ch_TD_NEGF}--\ref{Ch_Pert_leads} for the proofs and/or the rules for deriving other observables not included in this section.

The generic perturbation we consider takes the form
\begin{align}
\label{W}
\hat{\mathbf{W}}= \cos\omega t \sum_{nm} \mathbf{W}_{nm} c_{n}^{\dagger}c_{m}
\end{align} 
where $\mathbf{W}$ is a rather arbitrary (hermitian) matrix to be specified below.
This form of perturbation is suitable for AC electric field, AC magnetic field or lattice deformation.
\subsection{External AC perturbation}%
We first consider the situation where a uniform AC voltage is applied to one of the contacting electrodes,  say contact $m'$. The perturbation
takes the form
\begin{align}
\label{W_leads}
\mathbf{W}=eV_{ac} 1_{\bar m'}
\end{align} 
where $1_{\bar m'}$ is the identity matrix inside region ${\bar m'}$. The expressions below are given in the so-called wide-band limit (WBL)
where the energy dependance of the electrodes (retarded self-energies) is neglected. The most general (but also cumbersome) expressions without this approximation are given in Appendix \ref{Cookbook_general}.

The first important observable is the current $I_m(t)$ flowing through contact $m$. It can be decomposed according to the different harmonics of the perturbation,
\be
I_m(t)= \mathrm{Re} \sum_{l=0}^{\infty} I_m(l\omega) e^{-il\omega t}.
\ee 

The AC conductance matrix \cite{Buttiker_AC} (in the absence of DC bias) reads \cite{Guo_AC_conductance_NTs}:
\begin{align}
\label{AC_conductance_outside}
\begin{split}
\Upsilon_{m,m'}(\omega)\equiv \frac{dI_m(1\omega)}{dV_{ac}}=\frac{e^2}{h}\int dE\mathrm{Tr}\left[\Gamma_m\mathcal{G}_{2}\Gamma_{m'}\mathcal{G}^{\dagger}_{0}\right.\\
\left.-i\delta_{m,m'}\left(\mathcal{G}_2-\mathcal{G}^{\dagger}_0\right)\Gamma_m\right]\frac{f(E)-f(E+\hbar\omega)}{\hbar\omega}.
\end{split}
\end{align}
At small frequency, it simply reduces to the Landauer formula (\ref{landauer}).

In the WBL, we find an absence of rectification:
\be
\label{rectificationWBL}
\frac{d^2I_m(0\omega)}{dV_{ac}^2}= 0.
\ee
However this is not the case in general, as seen in Eq.(\ref{rectification}) in  Appendix \ref{Cookbook_general}. Let us briefly discuss what the above result implies on the modeling of the electrodes, i.e. on the effect of the position of the dashed line in Fig.\ref{System} (where the electrodes end and the system starts). Suppose that a real system is made of a small mesoscopic region connected to contacts. The contacts themselves are made of two parts: a very wide metallic part (energy independent, connected to the measuring apparatus) followed by a narrower region (energy dependent) which is itself connected to the mesoscopic region. From the modeling point of view, one must decide
where the electrodes start (in the wide or narrow region). At the quantum mechanical level, the position of the electrodes is totally arbitrary and simply corresponds to which degrees of freedom are integrated out. Therefore at this level, the physics is unaffected by the position where one chooses to define the electrodes  (in the wide region or in the narrower one). The fact that Eq.(\ref{rectification}) gives a non zero result (lead in the narrower region) or a vanishing one [lead in the wide region where Eq.(\ref{rectification}) reduces to Eq.(\ref{rectificationWBL})] therefore indicates that the difference between the two cases takes place at the statistical physics level, i.e. upon assuming that each electrode always remains at its thermal equilibrium. The correct choice between the two above mentioned possibilities depends on the inelastic mean free path: when almost no inelastic collisions take place in the narrower region, the electrodes should be considered to be in the wide region and the rectification effect vanishes. At higher temperature, the inelastic mean free path decreases, and the narrower region eventually becomes thermalized, which leads to a non zero rectification effect.

Another interesting limit is the adiabatic limit when the frequency $\omega$ is very small while the amplitude of the perturbation $V_{ac}$ can remain arbitrary large. To zeroth order in $\hbar\omega$, the current (for $m\neq m'$) is simply given by a trivial extension of the DC result,
\begin{align}
\label{Adiabatic_current}
I^{ad}_m(t)=\frac{e}{h}\int dE\left(-\frac{\partial f}{\partial E}\right)\mathrm{Tr}\left[\mathcal{G}_0\Gamma_{m'}\mathcal{G}^{\dagger}_0\Gamma_m\right]eV_{ac}\cos\omega t.
\end{align}
This expression is linear in $V_{ac}$ in the WBL. However, in general (see Eq.(\ref{adiabatic})) adiabatic current contains \textit{all} higher orders in amplitude as well. More interestingly, the first correction to adiabaticity reads,
\begin{align}
\begin{split}
\delta I^{ad}_m(t)=\frac{ie\omega}{4\pi}\int dE\left(-\frac{\partial f}{\partial E}\right)\mathrm{Tr}\left[\mathcal{G}_0\Gamma_{m'}\frac{\partial\mathcal{G}^{\dagger}_0}{\partial E}\Gamma_m\right.\\
\left.-\frac{\partial\mathcal{G}_0}{\partial E}\Gamma_{m'}\mathcal{G}^{\dagger}_0\Gamma_m\right]eV_{ac}\sin\omega t.
\end{split}
\end{align}
Note that while the adiabatic current follows exactly the slow changes of voltage, this correction is out of phase.

Another important observable is the electronic density $n(i,t)=\langle c^{\dagger}_i(t) c_i(t) \rangle$ on site $i$
whose decomposition in harmonics reads
\be
\label{Density_on_site}
n(i,t)= n_{eq}(i) + \mathrm{Re} \sum_{l=0}^{\infty} n(i,l\omega,m') e^{-il\omega t}
\ee 
where $n_{eq}(i)$ is the stationnary density in the absence of the time dependent potential. 
We refer to the response function $dn(i,1\omega,m')/dV_{ac}$ as the generalized injectivity.
It is a straightforward generalization of the injectivity defined in\cite{Buttiker_JOP1993,Buttiker_Nuovo_Cimento} at small frequency and it reads\cite{Wei_Wang_Curr_cons_NEGF_AC_Trans},
\begin{align}
\label{Injectivity}
\frac{dn(i,1\omega,m')}{dV_{ac}}=\frac{e}{h\omega}\int dE\left(f(E)-f(E+\hbar\omega)\right)\!\left[\mathcal{G}_2\Gamma_{m'}\mathcal{G}^{\dagger}_{0}\right]_{ii}
\end{align}

\subsection{Internal AC perturbation}%
Let us now turn to the case where the  AC perturbation is applied inside the device region,
\begin{align}
\label{W_device}
\mathbf{W}=eV_{ac} \mathbf{W}_{\bar 0\bar 0}
\end{align}
where the matrix block $\mathbf{W}_{\bar 0\bar 0}$ can take an arbitrary form inside the device region, allowing one to include many types of perturbation such as electric gates or time dependent magnetic field.

In the WBL the linear in $V_{ac}$ current response is \cite{Wei_Wang_Curr_cons_NEGF_AC_Trans}
\begin{align}
\label{AC_conductance_inside}
\frac{dI_m(1\omega)}{dV_{ac}}=\frac{ie^2}{h}\int dE\left(f(E)-f(E+\hbar\omega)\right)\mathrm{Tr}\left[\Gamma_m\mathcal{G}_{2}W\mathcal{G}^{\dagger}_{0}\right].
\end{align}

The DC (rectified) current contains only even orders in $V_{ac}$ (because sign of $V_{ac}$ is just a phase shift in Eq.(\ref{W})). Thus, the leading order contribution (in the absence of DC bias) reads \cite{Hanggi_PhysRep2005,Arrachea_Moskalets06},
\begin{align}
&\frac{d^2I_m(0\omega)}{dV_{ac}^2}=\frac{e^3}{4h}
\int dE\Biggl\{\notag\\
&\left(f(E)-f(E+\hbar\omega)\right)\mathrm{Tr}\left[\mathcal{G}_0W\mathcal{G}_2\Gamma\mathcal{G}^{\dagger}_2W\mathcal{G}^{\dagger}_0\Gamma_m\right]\\
&-\left(f(E-\hbar\omega)-f(E)\right)\mathrm{Tr}\left[\mathcal{G}_0W\mathcal{G}_{-2}\Gamma\mathcal{G}^{\dagger}_{-2}W\mathcal{G}^{\dagger}_0\Gamma_m\right]\Biggr\}\notag,
\end{align}
where $\Gamma=\sum_m\Gamma_m$.

Similarly, the $2^{\rm nd}$ harmonics of the current is given by,
\begin{align}
&\frac{d^2I_m(2\omega)}{dV_{ac}^2}=
\frac{ie^3}{2h}\int dE \Biggl\{\notag\\
&\left(f(E)-f(E+\hbar\omega)\right)\mathrm{Tr}\left[\mathcal{G}_2W\mathcal{G}^{\dagger}_0W\mathcal{G}^{\dagger}_{-2}\Gamma_m\right]\\
&+\left(f(E-\hbar\omega)-f(E)\right)\mathrm{Tr}\left[\mathcal{G}_2W\mathcal{G}_0W\mathcal{G}^{\dagger}_{-2}\Gamma_m\right]\Biggr\}.\notag
\end{align}

A particularly interesting case is the current generated upon perturbating the onsite potential on site $i$ (i.e. $\mathbf{W}_{kl}=eV_{ii} \delta_{ik}\delta_{il}$). The AC response function is the (generalized) emissivity\cite{Buttiker_JOP1993,Buttiker_Nuovo_Cimento} and it reads \cite{Wei_Wang_Curr_cons_NEGF_AC_Trans},
\begin{align}
\label{Emissivity}
\frac{dI_m(1\omega)}{dV_{ii}}=\frac{ie^2}{h}\int dE\left(f(E)-f(E+\hbar\omega)\right)\left[\mathcal{G}^{\dagger}_{0}\Gamma_m\mathcal{G}_{2}\right]_{ii}.
\end{align}
Note that the emissivity defined by B\"uttiker has a meaning of the number of particles (rather than the current) being emitted into the lead $m$ as a consequence of an internal perturbation. Thus, to come back to the original definition one has to multiply Eq.(\ref{Emissivity}) on both sides by $1/(ie\omega)$. Then its relation to the introduced above injectivity, Eq.(\ref{Injectivity}), will become obvious.

Last, we introduce the frequency-dependent Lindhard function\cite{Buttiker_JOP1993,Buttiker_Nuovo_Cimento} that relates a change of density on site $j$ to the perturbation on site $j'$ as \cite{Wei_Wang_Curr_cons_NEGF_AC_Trans},
\begin{align}
\label{Lindhard}
&\Pi(1\omega,j,j')\equiv\frac{dn(j,1\omega)}{dV_{j'j'}}\notag\\
&=-\frac{ie}{2\pi}\int dE\Bigl\{\left(f(E)-f(E+\hbar\omega)\right)\left[\mathcal{G}_2\right]_{jj'}\left[\mathcal{G}_0^{\dagger}\right]_{j'j}\\
&+f(E+\hbar\omega)\left[\mathcal{G}_2^{\dagger}\right]_{jj'}\left[\mathcal{G}_0^{\dagger}\right]_{j'j}-f(E)\left[\mathcal{G}_2\right]_{jj'}\left[\mathcal{G}_0\right]_{j'j}\Bigr\},\notag
\end{align}
where we implied an expansion similar to Eq.(\ref{Density_on_site}) for the electronic density on site $j$. We note that this expression does not assume the WBL. 

\subsection{Coulomb interactions: screening}\label{Screening}

The expressions given so far do not take electron-electron interaction into account. Besides missing potentially relevant
correlated physics, the non interacting expressions violate two important laws: conservation of current (some finite AC charge may pile up in the device region) and "gauge invariance" (raising the AC voltages of all leads simultaneously may produce some finite AC current),
\begin{align}
\sum_{m}\Upsilon_{m,m'}\ne 0 \ \ \ \sum_{m'}\Upsilon_{m,m'}\ne 0. \label{Curr_cons_GI}
\end{align} 
However, at a scale large enough, some screening will eventually take place to restore the global electric neutrality of the system (and these two laws). A minimum treatment of electron-electron interactions therefore implies solving the transport equations together with the Poisson equation,
following the general framework developed  in the group of B\"uttiker\cite{Buttiker_JOP1993,Buttiker_Nuovo_Cimento}. In practice, one has to 
solve the Poisson equation for the AC electric potential $V(\vec r,t)$ (we use spatial notations $\vec r$ instead of general coordinates $i$ in this subsection) which for the first AC harmonics reads,
\begin{align}
&\Delta \frac{dV(\vec r,1\omega,m')}{dV_{ac}}=-\frac{e}{\epsilon}\left[\frac{dn(\vec r,1\omega,m')}{dV_{ac}}\right.\notag\\
&\left.+\int d\vec r' \Pi(1\omega,\vec r,\vec r')\frac{dV(\vec r',1\omega,m')}{dV_{ac}} \right]
\end{align} 
($\epsilon$: dielectric constant). In a second step, one calculates the full AC current response in the leads as,
\be
\frac{dI_m(1\omega)}{dV_{ac}}=\Upsilon_{m,m'}(\omega)+  \int d\vec r \frac{dI_m(1\omega)}{dV({\vec r})} \frac{dV(\vec r,1\omega,m')}{dV_{ac}}
\ee
We therefore find that the full knowledge of the injectivity Eq.(\ref{Injectivity}), emissivity Eq.(\ref{Emissivity}) and Lindhard function Eq.(\ref{Lindhard}) is needed to complete a full self-consistent calculation\cite{Cai_SelfCons_AC_NEGF,Wei_Wang_Curr_cons_NEGF_AC_Trans,Kienle_SelfCons_AC_in_device_NEGF}.

\subsection{Relation to the scattering matrix formalism}\label{Ch_Relation_SM_formalism}
The Green's function approach taken here can be equivalently recast into the scattering approach developped in \cite{Buttiker_AC,Buttiker_PRL1993,Buttiker_PRL1993_1}. In particular, the scattering matrix $S(E)$ is connected to the retarded Green's function $\mathcal{G}(E)$ through the Fisher-Lee relation \cite{Fisher_Lee,Datta1997}. In the WBL, the equivalent expressions in the scattering formalism are simply obtained by using the \textit{formal} substitution,
\begin{align}
\sqrt{\Gamma_n}\mathcal{G}_0\sqrt{\Gamma_m}\rightarrow -i S_{nm}(E) -i\mathbf{1}
\end{align}
where $S_{nm}(E)$ is the scattering submatrix between the electrodes $m$ and $n$. For instance, for the AC conductance, Eq.(\ref{AC_conductance_outside}), the substitution provides,
\begin{align}
\label{Buttiker_AC_conductance}
\Upsilon_{m,m'}(\omega)&=-\frac{e^2}{h}\int\frac{dE}{\hbar\omega}\left(f(E)-f(E+\hbar\omega)\right)\notag\\
&\times\mathrm{Tr} [\delta_{mm'} -  S^{\dagger}_{mm'}(E)S_{mm'}(E+\hbar\omega) ],
\end{align}
which is precisely the expression derived in Ref.\cite{Buttiker_AC}. Note that the trace in Eq.(\ref{Buttiker_AC_conductance}) is taken over the open channels in the lead.

In general (not in the WBL) mapping between the two formalisms requires solving the scattering problem in presence of the oscillating field via the Floquet approach. For more details on this approach and its relation to the NEGF see Ref.\cite{Arrachea_Moskalets06}.

\section{Applications}\label{Ch_Applications}

In this section, we apply the formalism on three practical examples: the AC conductance of a simple one dimensional chain, a quantum point contact (QPC) and an electronic Mach-Zehnder (MZ)  interferometer in the quantum Hall regime. These examples, of increasing complexity, are chosen to illustrate how the numerical calculations can be performed in practice and how the AC physics can provide insights absent in DC. The stationary Green's functions $\mathcal{G}_{l}$ at the root of the AC expressions were obtained with the knitting algorithm described in Ref.\cite{Waintal_KNIT}.

The AC observable we consider is the AC conductance, which is given by Eq.(\ref{AC_conductance_outside_general}). 
Eq.(\ref{AC_conductance_outside_general}) can be written as the sum of three terms,
\begin{align}
\label{Divergency_integrands}
&\Upsilon_{21}(\omega)=-\frac{e^2}{h}\int \frac{dE}{\hbar\omega}\mathrm{Tr}\left[\left(f(E)-f(E+\hbar\omega)\right)A_{21}^{ar}\right.\notag\\
&\left.-f(E)A_{21}^{rr}+f(E+\hbar\omega)A_{21}^{aa}\right],
\end{align}
with
\begin{align}
\label{AR}
A^{ar}_{21}&=\mathrm{Tr}\left[\Lambda^{ar}_2(E;E+\hbar\omega)\mathcal{G}_{2}\Lambda^{ra}_{1}(E+\hbar\omega;E)\mathcal{G}^{\dagger}_{0}\right]\\
\label{RR}
A^{rr}_{21}&=\mathrm{Tr}\left[\Lambda^{rr}_2(E;E+\hbar\omega)\mathcal{G}_{2}\Lambda^{rr}_{1}(E+\hbar\omega;E)\mathcal{G}_{0}\right]\\
\label{AA}
A^{aa}_{21}&=\mathrm{Tr}\left[\Lambda^{aa}_2(E;E+\hbar\omega)\mathcal{G}^{\dagger}_{2}\Lambda^{aa}_{1}(E+\hbar\omega;E)\mathcal{G}^{\dagger}_{0}\right]
\end{align}
where
\begin{equation}
\label{Delta_x_x}
\Lambda^{cb}_m(E;E')=\mathsf{\Sigma}^c(m;E)-\mathsf{\Sigma}^{b}(m;E').
\end{equation}
(here $c,b\in \{ a,r\}$ stands for the retarded or advanced self energy). The three terms (\ref{AR}), (\ref{RR}) and (\ref{AA}) are direct outputs of recursive Green's function like techniques so that the main numerical difficulty lies in the evaluation of the integral over energies.

\subsection{Technical details on the numerical integration}\label{Ch_Technical_details_num}
%
\begin{figure}[]
\begin{center}
\includegraphics[angle=0,width=0.9\linewidth]{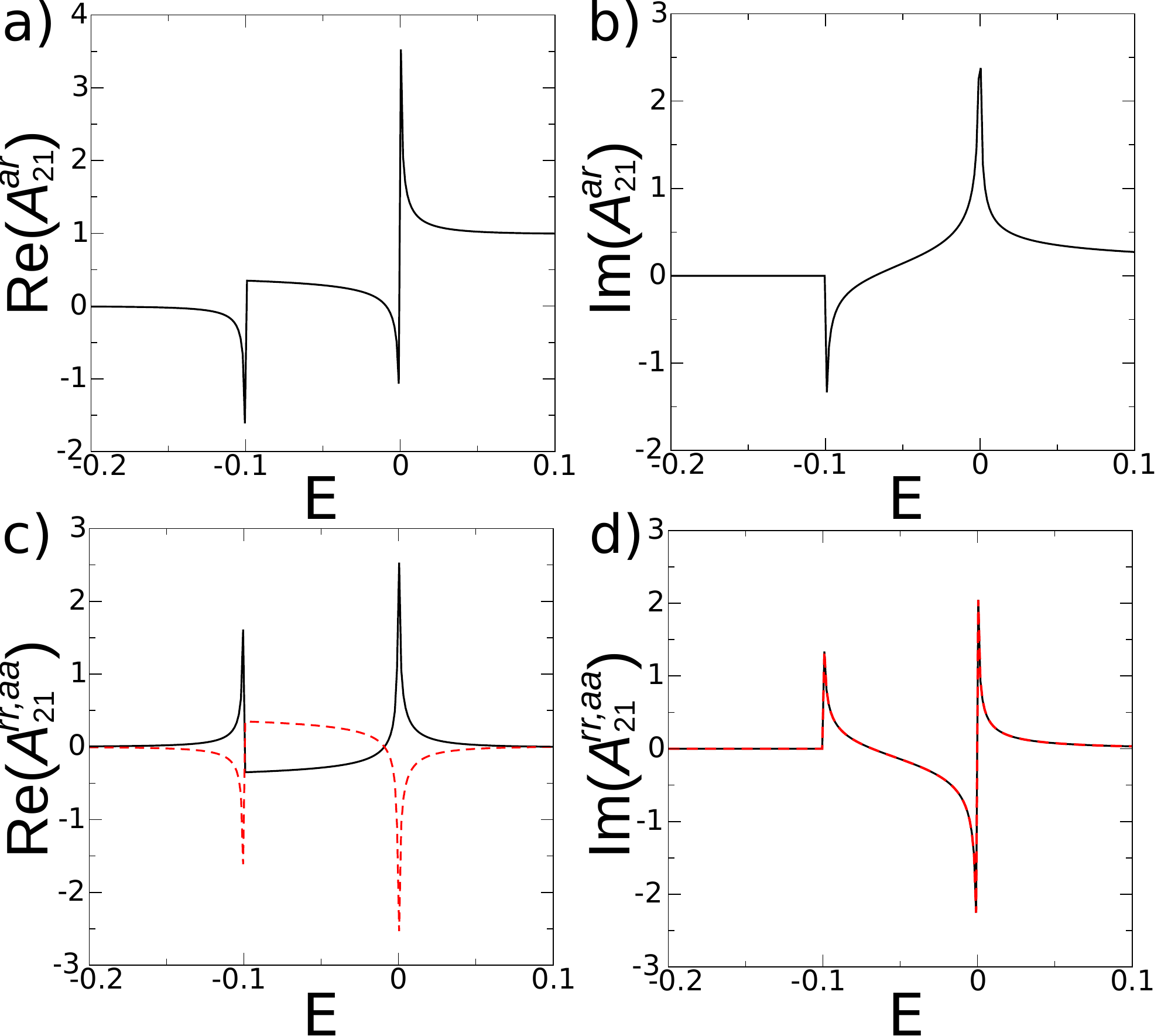}
\caption{(Color online): Real (left panels) and imaginary (right panels) parts of the integrands appearing in Eq.(\ref{Divergency_integrands}) as a function of energy close to the band edge $E=0$. The frequency was chosen to be $\hbar\omega=0.1$. There are pronounced peaks at energies $E=0$ and $E=-\hbar\omega$. Upper panels: $A_{21}^{ar}$. Lower panels: $A_{21}^{rr}$ (black solid line) and $A_{21}^{aa}$ (red dashed line).}
\label{1D_divergencies}
\end{center}
\end{figure}
We start with the AC conductance of a simple one dimensional chain described by the Hamiltonian,
\begin{align}
\label{1D}
H=2-\sum_{n=-\infty}^{\infty} \left(c^{\dagger}_{n+1}c_n + h.c.\right)
\end{align}
(the constant $2$ serves to offset the bottom of the band to $E=0$). The device region is of size $L$ so that we suppose that
the system stays in thermal equilibrium for $n\le 0$ (left lead, region $\bar 1$) and $n>L$ (right lead, region $\bar 2$).

For a single site $L=1$ device, the (onsite) Green's function of the system can be easily obtained \cite{Economou},
\begin{equation}
\mathcal{G}(E)=\left\{
\begin{array}{l}
-\frac{1}{\sqrt{(E-2)^2-4}}, \mbox{for} E\leq 0,\\
\frac{1}{i\sqrt{4-(E-2)^2}}, \mbox{for} |E|<4,\\
\frac{1}{\sqrt{(E-2)^2-4}}, \mbox{for} E\geq 4.
\end{array}
\right.
\end{equation}
We find that it contains square root singularities $1/\sqrt{E-E_0}$ which appear on the edges of the band (or more generally in quasi 1D system, whenever there is an opening/closing of a new conducting channel). Typical plots of the integrands are shown in Fig.~\ref{1D_divergencies}. These singularities are integrable but may require a very fine discretization mesh. In practice, we find that advanced integration routines, such as QUAD which is used in this work\cite{QUAD} (based on an adaptative mesh and Gauss quadrature formula), can handle these singularities properly and provide precise results within a reasonable computing time. A faster convergence is often obtained when one provides the routine with a precise location of the position of the singularities. This location can be found by simple calculations of the DC transmission of perfect wires as a function of energy using a dichotomy algorithm (the singularities occur at the energy where the transmission has a step like increase). An alternative route is to remove the singularities using a local change of variable. One starts by locating the singularities and dividing the integration range in small segments with (at most) one singularity on one boundary of the segment. The integration $\int_{E_0}^a dE f(E)$ on one segment (all those integrations can be done in parallel) is then performed using the local change of variable  $\bar E =\sqrt{E-E_0}$ which removes the singularity, i.e. one integrates $\int_0^{\sqrt{a-E_0}} 2 d\bar E g(\bar E)$ with the non diverging function $g(\bar E)\equiv\bar E f(\bar E^2 +E_0)$ (See also\cite{NumRecipes2007} and the straightforward improvement to include two singularities in one segment). 

\begin{figure}[]
\begin{center}
\includegraphics[angle=0,width=0.9\linewidth]{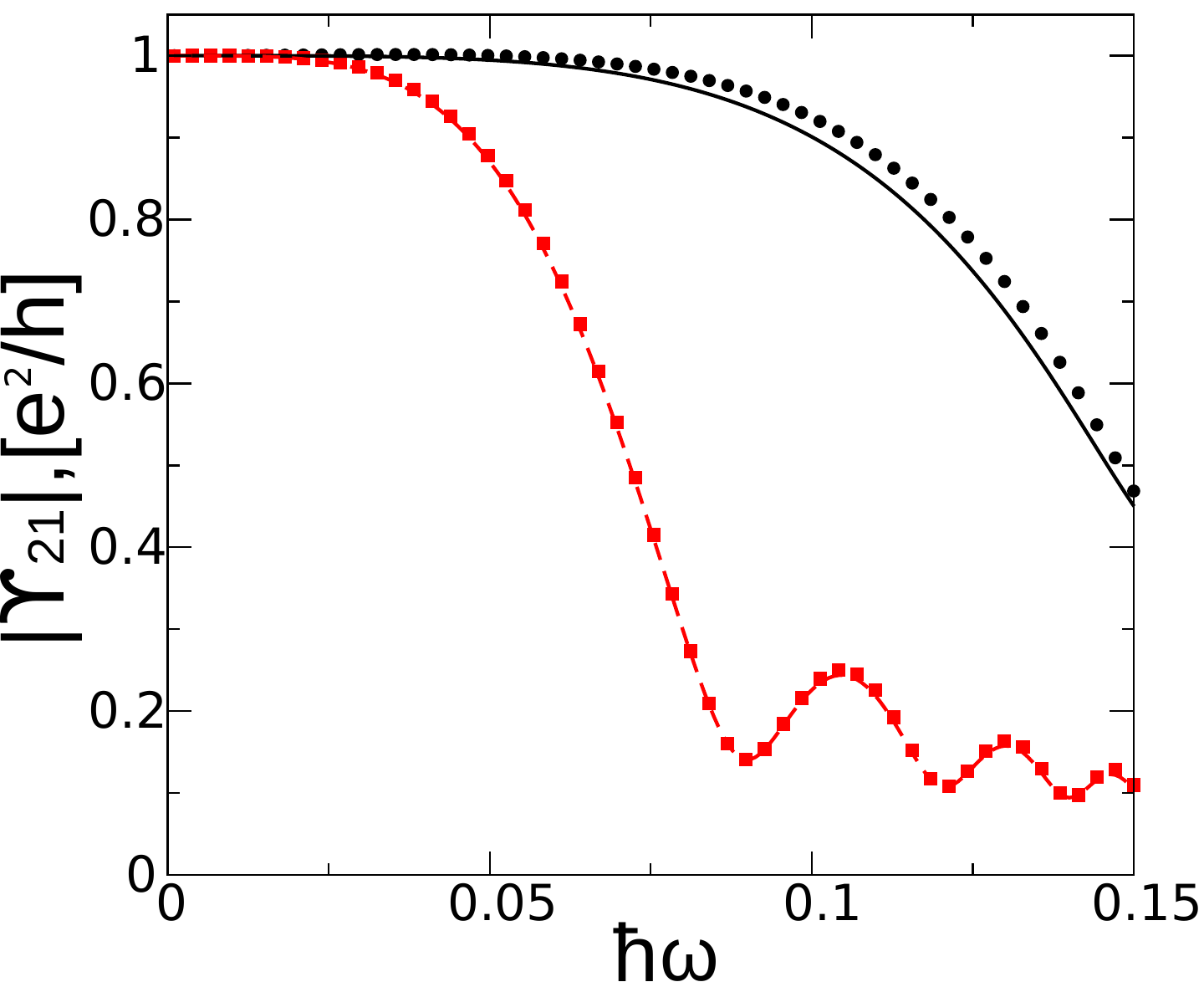}
\caption{(Color online): Absolute value of the AC left-to-right conductance for the one-dimensional wire of length $L=40$ (black cirles) and $L=200$ (red rectangles). Symbols: numerical calculation with the Green's function formalism, Eq.(\ref{Divergency_integrands}) (keeping only the $A_{21}^{ar}$ term), lines: scattering approach Eq.(\ref{Buttiker_AC_conductance}) (see text for details). We find a visible difference between the two approaches for the small size $L=40$ that disappears when $L$ increases. Fermi energy is $E_F=0.17$.}
\label{Buttiker_vs_GF}
\end{center}
\end{figure}

The AC conductance of the one dimensional wire is shown in Fig.~\ref{Buttiker_vs_GF} for two lengths $L=40$ and $L=200$. We find that the calculation performed keeping only the "WBL-like" term $A^{ar}_{21}$ (this is the only surviving term in the WBL) in Eq.(\ref{Divergency_integrands}), is equivalent to using Eq.(\ref{Buttiker_AC_conductance}), derived within the scattering approach\cite{Buttiker_AC}, in the large $L$ limit. In order to compare the two approaches, we integrated numerically (assuming zero temperature) Eq.(\ref{Buttiker_AC_conductance}) using the actual dispersion relation of the Hamiltonian (\ref{1D}), $E(k)=2-2\cos k$, and $S_{21}=\exp(ikL)$ (perfect transmission). To understand why both formalisms coincide when $L\rightarrow\infty$, note that terms $A^{rr}_{21}$ and $A^{aa}_{21}$ typically oscillate as $\exp\left[\pm i\left(k(E+\hbar\omega)+k(E)\right)L\right]\propto 
\exp\left[\pm 2i k_F L\right]$ ($k_F$: Fermi momentum) so their integrals quickly vanish when $k_F L \gg 1$ and only the (WBL) term $A^{ar}_{21}\propto\exp\left[i\left(k(E+\hbar\omega)-k(E)\right)L\right]$ remains (in agreement with Refs. \cite{Buttiker_AC,Buttiker_Blanter} where the fast oscillating terms are neglected in the derivation of the current operator).
%

\subsection{Quantum Point Contact}\label{QPC}

\begin{figure}[]
\begin{center}
\includegraphics[angle=0,width=0.9\linewidth]{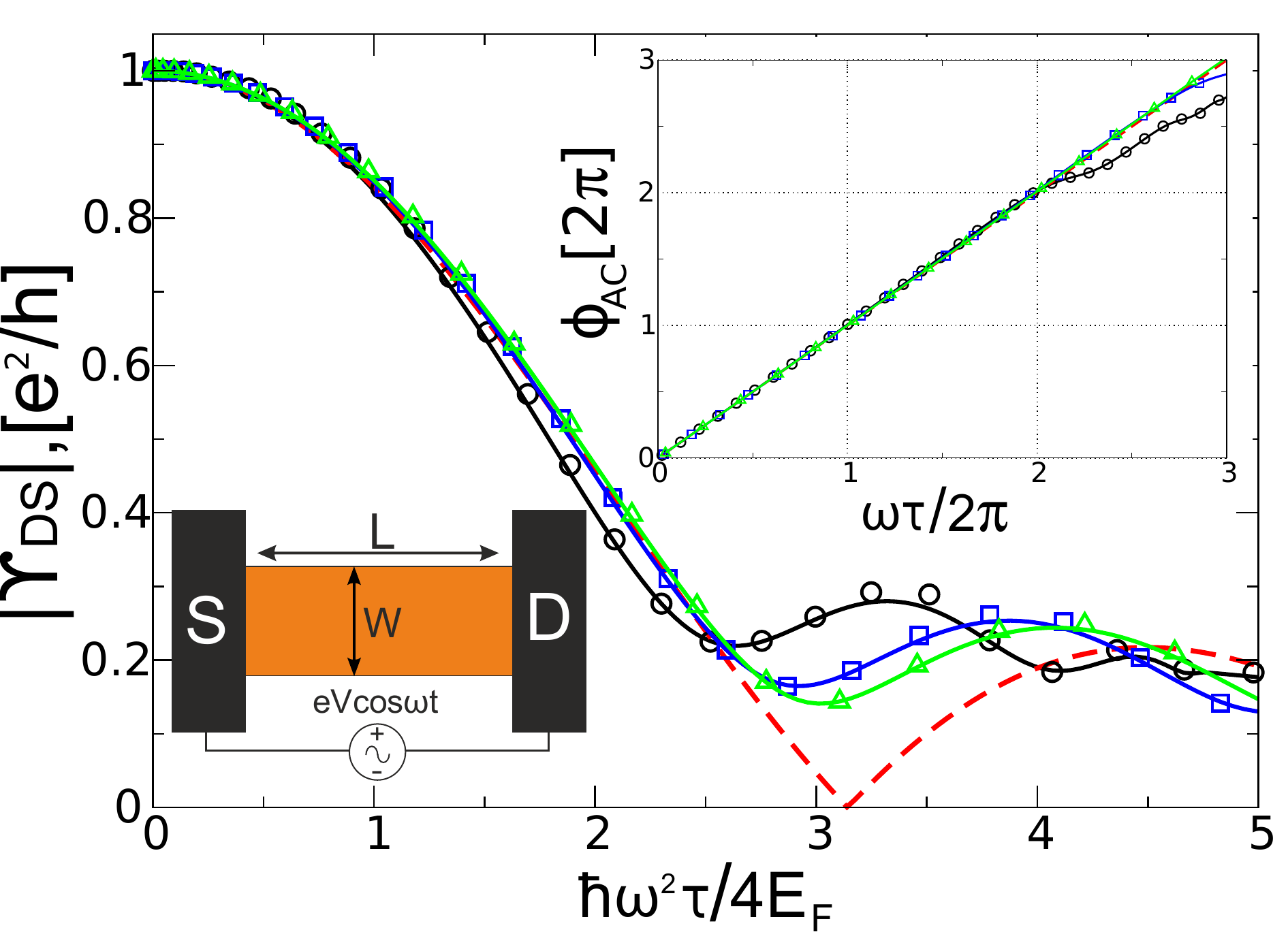}
\caption{(Color online): Rescaled amplitude and phase (upper inset) of the two terminal AC conductance $\Upsilon_{DS}=|\Upsilon_{DS}|e^{i\phi_{AC}}$ for a quasi one-dimensional wire of width $W=10$. Lower inset: schematic picture of the setup. Fermi energy is chosen to have only one propagating mode. The parameters are: $E_F=0.68\epsilon_2$, $L=100$ (black circles), $E_F=0.79\epsilon_2$, $L=150$ (blue rectangles), $E_F=0.79\epsilon_2$, $L=200$ (green triangles), where $\epsilon_2$ is the energy of the second mode opening (see Eq.(\ref{Disp_rel_q1D})). Different symbols correspond to the numerical integration of Eq.(\ref{Divergency_integrands}), while the lines are calculated with the help of Eq.(\ref{q1D_AC_conductance}) exploiting the full tight-binding dispersion relation (\ref{Disp_rel_q1D}). Red dashed line is the analytical fit using Eq.(\ref{G_AC_2DEG}). All the lengths are in units of the lattice constant.}
\label{Fig1}
\end{center}
\end{figure}

We continue with a quasi one-dimensional wire of width $W$ and length $L$ connected to two reservoirs. AC bias is applied to the source lead (S) and we are interested in the current response in the drain (D), see~Fig.~\ref{Fig1}. The Hamiltonian is the direct extension of Eq.(\ref{1D}) to the quasi 1D geometry.
\begin{align}
\label{quasi1D}
H=2-\sum_{n=-\infty}^{\infty} \sum_{m=1}^W \left(c^{\dagger}_{n+1,m} c_{n,m} + c^{\dagger}_{n,m+1} c_{n,m} + h.c.\right)
\end{align} 
The dispersion relation for this discrete model is (in units of the hopping constant)
\begin{equation}
\label{Disp_rel_q1D}
E_n(k)=\epsilon_n+2-2\cos k,
\end{equation}
wich corresponds,  in the continuum limit, $k\rightarrow 0$, to
\begin{equation}
\label{Disp_rel_q1D_cont}
E_n(k)=\epsilon_n+k^2,
\end{equation}
where the transverse energy $\epsilon_n=2-2\cos\left(n\pi/(W+1)\right),n=1\dots W$, and $k$ is the longitudinal momentum. The integer $n$ defines the quantized values of the transverse momentum and thereby enumerates the conducting channels.

We focus on the regime where only the first ($n=1$) channel is open and use the corresponding transverse energy $\epsilon_1$ as our reference energy. In this limit the system is described by a unique velocity, hence the time of flight through the system is well defined. In the scattering matrix approach, the system is described by its transmission matrix $S_{DS}(E)=\exp(ikL)$: a wave packet is entirely transmitted and only acquires a (energy dependent) phase. Then, equation (\ref{Buttiker_AC_conductance}) reads
\begin{align}
\label{q1D_AC_conductance}
\Upsilon_{DS}=\frac{e^2}{h}\int_{E_F-\hbar\omega}^{E_F}\frac{dE}{\hbar\omega}\exp\left\{iL\left[k(E+\hbar\omega)-k(E)\right]\right\}.
\end{align}
In the continuum limit, when $\hbar\omega\ll E_F$ it can be further simplified using Eq.(\ref{Disp_rel_q1D_cont}) into
\be
\Upsilon_{DS}=\frac{e^2}{h}e^{i\omega\tau}\frac{\sin{\left(\frac{\hbar\omega^2\tau}{4E_F}\right)}}{\left(\frac{\hbar\omega^2\tau}{4E_F}\right)},\label{G_AC_2DEG}
\ee
where $\tau=L/v_F$ is the time of flight from the source lead to the drain (see lower inset on Fig.~\ref{Fig1}). From this simple calculation we notice that the AC conductance gives access to two characteristic time (or energy) scales of the system. Indeed, the numerical results of Fig.~\ref{Fig1} indicate that the absolute value of the AC conductance and its phase can be very well described by this simple scaling law up to moderately large frequencies. The scaling parameters arising in this case allow for the extraction of the time of flight and the longitudinal Fermi energy. We note that, as the Fermi energy $E_F$ and thus the velocity $v_F$ that enter the previous expression are counted from $\epsilon_1$ (i.e. they are in fact the longitudinal Fermi energy and velocity), one can actually slow down the electrons to bring these times and energy scales into an experimentally accessible window (GHz range).

\begin{figure}[]
\begin{center}
\includegraphics[angle=0,width=1.0\linewidth]{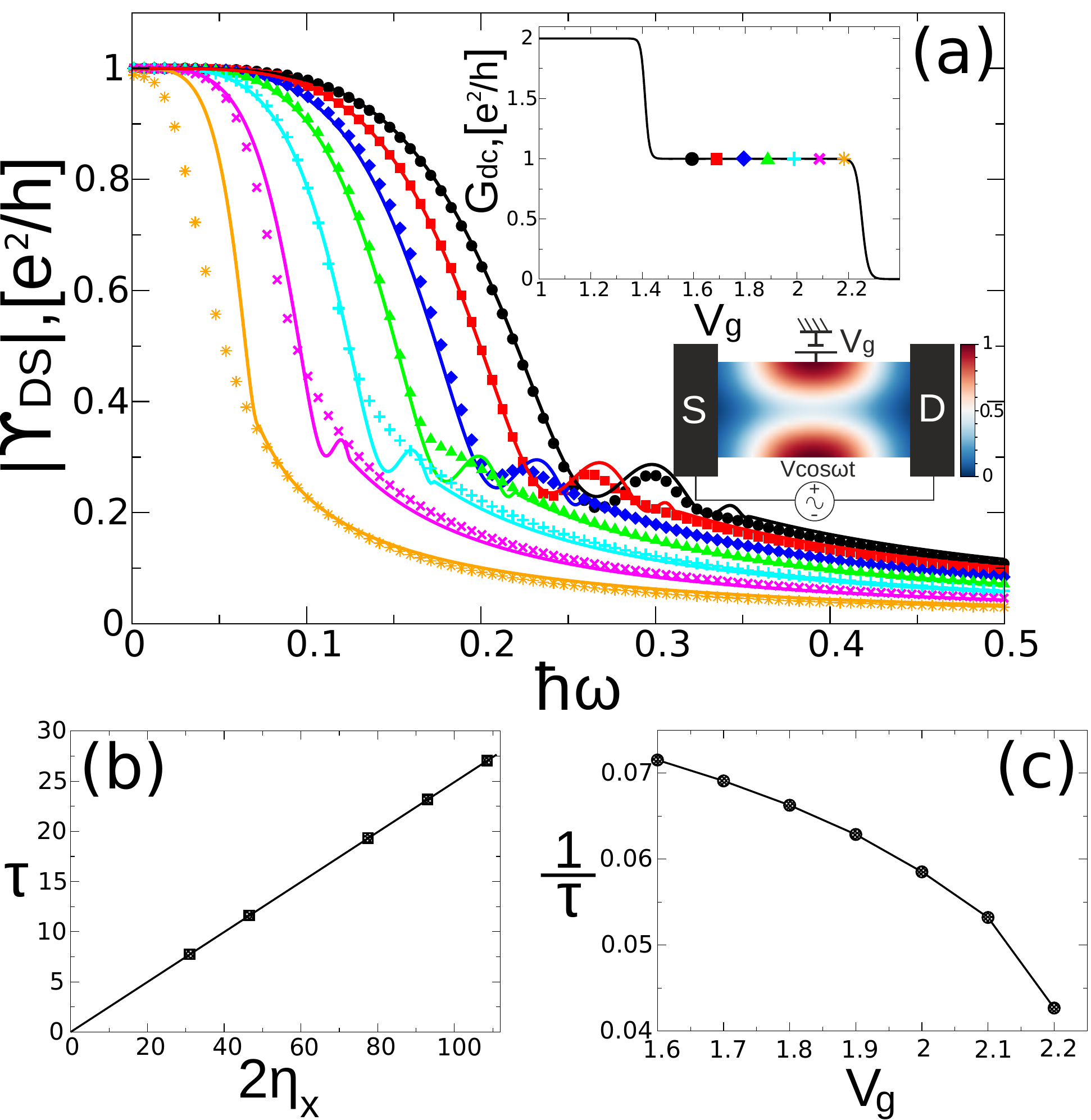}
\caption{(Color online): Quantum Point Contact. (a) Absolute value of the AC conductance as a function of driving frequency for different values of $V_g$ depicted on the upper inset. Upper inset: dc conductance as a function of $V_g$. The lower inset represents actual potential profile in the device forming the QPC. (b),(c) Extracted time of flight as a function of QPC length, $2\eta_x$, (see Eq.(\ref{V_QPC})) and $V_g$, respectively. Fermi energy in the calculations was $E_F=1.2$ providing 4 open channels in the leads, while only one being transmitted through the QPC. Other parameters: $W=11$, $L=124$, $\xi_x=\eta_x/2$, $\xi_y=3\eta_y/2$, and $\eta_y=2$.}
\label{Fig2}
\end{center}
\end{figure}

A practical way to implement an effective quasi-one dimensional wire is through a quantum point contact (QPC) formed by confining an electron gas with electrostatic gates placed on top of a semiconducting heterostucture. We add an electric potential to our quasi one dimensional wire Hamiltonian,
\begin{align}
\label{pot}
V=V_g \sum_{n=-\infty}^{\infty} \sum_{m=1}^W  \Phi_x(n-n_0)\Phi_y(m-m_0) c^{\dagger}_{n,m} c_{n,m} 
\end{align}
where we have used the following confining "saddle point" potential (see the lower inset of Fig.~\ref{Fig2}a for a color plot of the potential),
\begin{align}
\label{V_QPC}
\Phi_x(n)=&\frac{1}{2}\!\left[\tanh\left(\frac{n+\eta_x}{\xi_x}\right)+
\tanh\left(-\frac{n-\eta_x}{\xi_x}\right)\right]\\
\Phi_y(m)=&\frac{1}{2}\!\left[2\!-\!\tanh\left(\frac{m+\eta_y}{\xi_y}\right)\!-\!\tanh\left(\!-\frac{m-\eta_y}{\xi_y}\right)\right]
\end{align}
where the parameters $\xi_{x},\xi_y$ control the steepness of the potential (they are chosen big enough so that the potential rises essentially in an adiabatic way), and ($x_0$,$y_0$) determines the position of the center of the QPC. The effective length (width) of the QPC is $2\eta_{x}$ ($2\eta_y$). In this case the dispersion relation in the gated region may be considered similar to Eq.(\ref{q1D_AC_conductance}) except that now the transverse energy $\epsilon_n$ is determined not only by the width, but also depends on the parameters of the QPC ($\eta_y, \xi_y, V_g$, etc.). In our calculations we controlled the value of $V_g$ (keeping other parameters fixed) to have only one open channel (at a given $E_F$) in the gated region (see upper inset on Fig.~\ref{Fig2}a). The results of the numerical simulations of the AC conductance for this system are given in Fig.~\ref{Fig2}.
Fig.~\ref{Fig2}a shows the absolute value of the AC conductance as a function of the driving frequency. Different symbols correspond to the numerical results [Eq.(\ref{Divergency_integrands})] for different values of the gate voltage $V_g$, hence for different effective longitudinal velocities, as shown in the upper inset. The fitting lines are obtained from the scattering matrix formalism, Eq.(\ref{q1D_AC_conductance}), making use of the dispersion relation (\ref{Disp_rel_q1D}). Again, the transverse energy of the open mode, $\epsilon_1(\eta_{y},\xi_{y},V_g,...)$, was chosen as the energy reference. Comparing the two approaches, we see that the closer we are to the edge of the propagating mode (various symbols on the upper inset of Fig.~\ref{Fig2}a), the worse is the fit given by Eq.(\ref{q1D_AC_conductance}). This is due to the fact that the scattering matrix formula is applicable when the Fermi velocity is a smooth slowly varying function of energy\cite{Buttiker_Blanter}, which  breaks down near the band edge ($E_F\approx\epsilon_1$). 

From the slope of the phase of the AC conductance ($\phi=\omega\tau$, curves similar to the inset of Fig.~\ref{Fig1}, not shown) we can extract the effective time of flight $\tau$ of the electrons through the QPC, see Figs.~\ref{Fig2}b,c. We find that $\tau$ scales linearly with the QPC length $2\eta_x$ (ballistic transport) while it increases as we tune $V_g$ towards the closing of the propagating mode (the velocity vanishes when the mode becomes evanescent, $E_F<\epsilon_1$). The above calculations did not take screening into account. However, as all the calculations were performed on perfectly transmitting channels there is no "Landauer dipole" in the problem.
Calculations done assuming perfect screening (enforcing current and gauge conservation\cite{Buttiker_PRL1993,Aronov_frequency_dependence_AC_QPC} suggest that the above physical signatures remain observable in presence of screening, and in particular should allow for the measurement of the time of flight.

\subsection{Mach-Zehnder interferometer}\label{MZ}
\begin{figure}[!htb]
\begin{center}
\includegraphics[angle=0,width=1.0\linewidth]{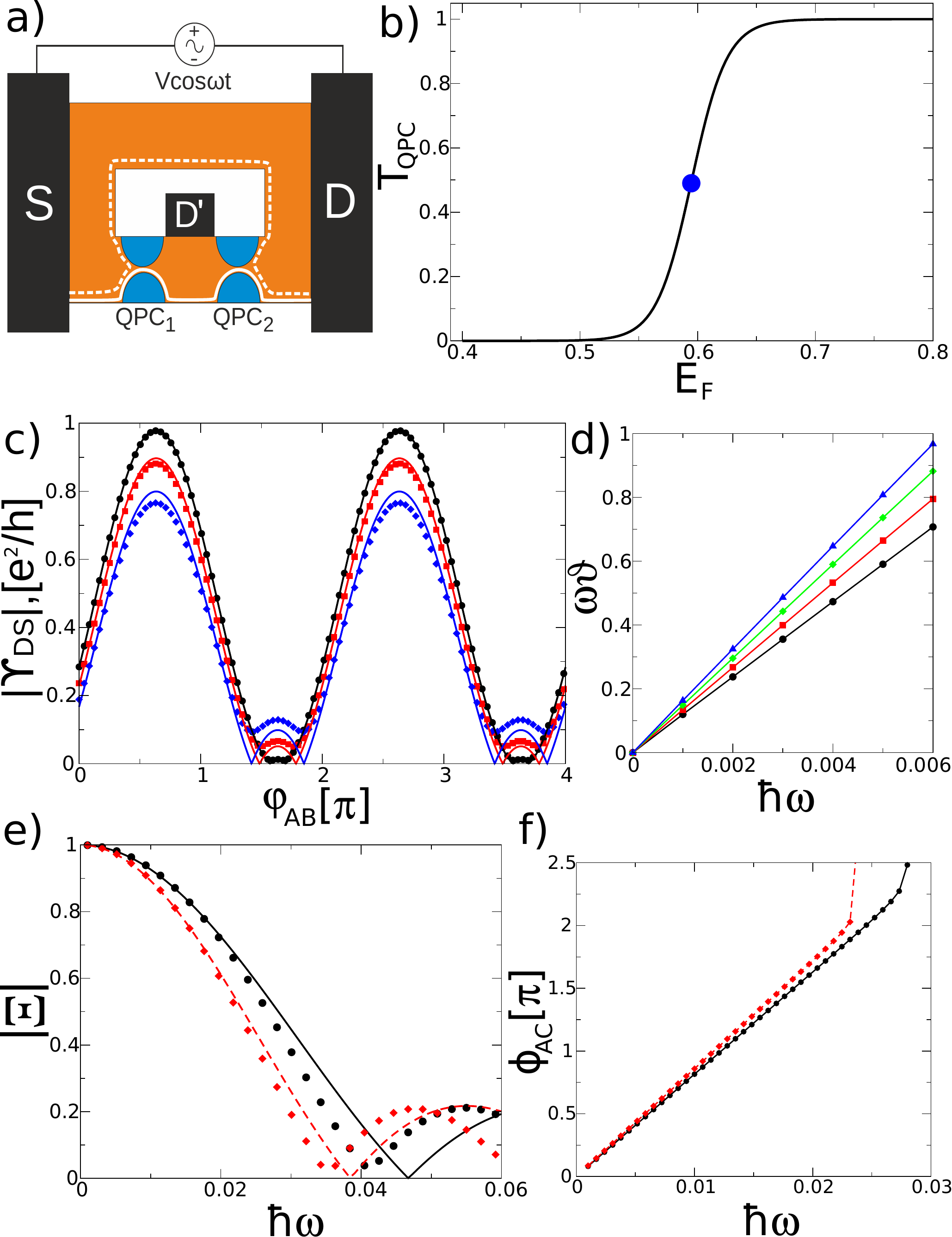}
\caption{(Color online): Mach-Zender interferometer. (a) Cartoon of the system with the interfering paths represented by the solid and dashed white lines. Length of the sample $L=80$, width of the system $W=W_h+22$, where $W_h$ is the width of the hole. The system is in the quantum Hall regime at filling factor $\nu=1$. The two QPCs were chosen to be semi-transparent  $T_{1,2}=\frac{1}{2}$. (b) Transmission characteristics of the QPCs. The blue dot corresponds to the Fermi energy at which QPC is half transparent. (c) AB oscillations of AC conductance as a function of magnetic flux through the hole. Black circles, red rectangles, and blue diamonds correspond to the driving frequency $\hbar\omega=0.003,0.007,0.01$ respectively. Width of the hole $W_h=45$. (d) $\omega\vartheta$ as a function of frequency for different values of $W_h$ extracted with Eq.(\ref{G_AC_MZ_FINAL}) (see text for details). From bottom up $W_h=30,35,40,45$. (e) Function $|\Xi(\omega)|$, see Eq.(\ref{Xi}), as a function of frequency. Black circles and red diamonds correspond to $W_h=35,45$ respectively. (f) Respective phase of the AC conductance $\phi_{AC}=\omega\tau$ as a function of frequency. In plots (c) and (e) all the symbols were calculated with the Green's function formalism, Eq.(\ref{Divergency_integrands}), modeling the QPCs with Eq.(\ref{V_QPC}). The connecting lines are the corresponding analytical fits, Eq.(\ref{G_AC_MZ_FINAL}), with parameters calculated from the Green's function-based numerics.}
\label{Fig3}
\end{center}
\end{figure}

We close this section with a discussion of the AC response of an electronic MZ interferometer in the quantum Hall regime \cite{Ji2003Nat,Heiblum_PRL_MZ,Heiblum_PRL07,Heiblum_NatPhys07,Roulleau_PRB07,Roulleau_PRL08,Roulleau_PRL08bis,Roulleau_PRL09}. The setup consists of the two-dimensional electron gas confined in a finite geometry, connected to three reservoirs: source (S), drain (D) and internal drain (D'). Fig.~\ref{Fig3}a presents the sample together with a schematic of the two interfering edge states.
The corresponding Hamiltonian of the system is
\begin{align}
\label{HMac}
H=2-\sum_{n,m \in S} e^{im\phi} c^{\dagger}_{n+1,m} c_{n,m} + &c^{\dagger}_{n,m+1} c_{n,m} + h.c.\nonumber \\
+&V_{n,m} c^{\dagger}_{n,m} c_{n,m}
\end{align} 
where $\phi$ is a Peierls phase\cite{Datta1997} in the Landau gauge (magnetic flux through a lattice plaquette in units of $e/\hbar$) and $S$ the list of sites in the system (orange region
in Fig.~\ref{Fig3}a plus the semi infinite leads). The system is put in a strong magnetic field driving the system into the quantum Hall regime. We focused on the case with a filling factor $\nu=1$ (single edge channel). The additional electrode (D') is necessary to avoid the paths to make multiple loops (that would turn the interferometer into a Fabry-Perot). The electric potential $V_{n,m}$ defines the two QPCs  in the lower arm of the interferometer (see the previous section). The two QPCs play the role of  beam splitters for the quantum Hall edge state. The interfering paths are shown by the solid and dashed lines on Fig.~\ref{Fig3}a. We put an additional Aharonov-Bohm~(AB) flux $\varphi_{AB}$ through the hole of the interferometer that allows to change the relative phase between the paths without changing the edge states. We calculated the AC conductance as a function of the AB flux $\varphi_{AB}$ and frequency $\omega$ as shown in Fig.~\ref{Fig3}. 
A similar setup was studied in DC in Ref.\cite{Waintal_KNIT}.

We start with simple analytical considerations using the scattering matrix approach for the edge states. Let us assume, for simplicity, that both QPCs are characterized by energy independent transmission probabilities $T_{1,2}$ (and corresponding reflection probabilities are $R_{1,2}=1-T_{1,2}$). We describe the QPCs by their scattering matrices, which can be parametrized as follows
\begin{equation}
\left( \begin{array}{cc}
i\sqrt{R_k} & \sqrt{T_k} \\
\sqrt{T_k} & i\sqrt{R_k} \end{array} \right), \: k=1,2.
\end{equation}
The source-to-drain transmission amplitude $S_{DS}$ then consists of the contributions from two paths, path $a$ (solid line on Fig.~\ref{Fig3}a) which is a consequence of two sequential transmissions through the QPCs and path $b$ (dashed line) arising from sequential reflections
\begin{align}
\label{T_DS}
S_{DS}(E)=\sqrt{T_1T_2}e^{i\psi_a(E)}-\sqrt{R_1R_2}e^{i\psi_b(E)}.
\end{align}
Traversing either path, an electron acquires a phase $\psi_{a,b}$. The phase itself contains two contributions, the AB phase caused by the magnetic flux through the hole and a dynamical phase of the propagating plain wave along the path,
\begin{align}
\label{Phase_AB}
&\psi_{x}(E)=k(E)L_{x}+\varphi_{x},\; x=\{a,b\},\\
&\varphi_{AB}=\varphi_{b}-\varphi_{a},
\end{align}
where $k$ is the longitudinal wave number of the edge state and $L_{x}$ is a length of the corresponding path. In order to calculate the AC conductance we need to specify the energy dependence of the phase factors in Eq.(\ref{T_DS}). We note that by varying energy we modify only the dynamical part of the phase, while the AB flux is unaffected. Thus, we make the following ansatz
\begin{align}
\label{Phase_F_of_E}
&\psi_{x}(E)\approx\psi_{x}(E_F)+\lambda_{x}(E-E_F),\\
&\lambda_{x}=L_{x}\frac{\partial k}{\partial E}(E_F),\;x=\{a,b\}.
\end{align}
Actual value of $\psi_x(E_F)$ and $\lambda_x$ depends on the boundary conditions defining the geometry of the setup. Assuming we know the edge state dispersion relation, we can relate the latter to the group velocity of the edge state $v_g$ at the Fermi level via $\lambda_{x}=L_{x}/(\hbar v_g)$. 

At this stage we are prepared to calculate the AC conductance with equations (\ref{Buttiker_AC_conductance}), (\ref{T_DS}), and (\ref{Phase_F_of_E}). The scattering formula is valid when $\hbar\omega\ll E_F$ and we will use it to carry out the integration in energy. Let's assume, for simplicity, that the QPCs are tuned to half transparency $T_{1,2}=1/2$. Then, after a straightforward calculation we obtain
\begin{align}
\label{G_AC_MZ_FINAL}
\Upsilon_{DS}(\omega,\varphi_{AB})&=\frac{1}{2}e^{i\omega\tau}\left[\cos\frac{\omega\vartheta}{2}\right.\notag\\
&\left.-\frac{\sin\frac{\omega\vartheta}{2}}{\frac{\omega\vartheta}{2}}\cos\left(\psi_b(E_F)-\psi_a(E_F)\right)\right],\\
&\tau=\frac{L_b+L_a}{2v_g},\:\vartheta=\frac{L_b-L_a}{v_g}.\label{Tau_Theta}
\end{align}
We have two time scales naturally appeared, namely the average time of flight $\tau$ and the relative time $\vartheta$, coming from the asymmetry between the paths.

Now we turn to our numerical results (see Fig.~\ref{Fig3}) and compare them to the simple formula (\ref{G_AC_MZ_FINAL}). On Fig.~\ref{Fig3}b we plot the DC characteristics of the QPCs considered in our modelling. The Fermi level is fixed at a half transparency value. We remind that in order to obtain Eq.(\ref{G_AC_MZ_FINAL}) we assumed that the transmission characteristics of the QPCs are energy independent. However, as one can see from Fig.~\ref{Fig3}b, in our sample detuning from the assumed value becomes important for $\hbar\omega\simeq 0.01$ and is of the order $\Delta T_{1,2}\simeq 0.15$. Next, Fig.~\ref{Fig3}c represents the plots of the AB oscillations of the AC conductance as a function of flux for three different values of driving frequency. Symbols of different types on the Figure represent the Green's function based calculation, Eq.(\ref{Divergency_integrands}), while the solid lines are corresponding fits according to Eq.(\ref{G_AC_MZ_FINAL}). This fit is obtained as follows: (i) we perform a DC calculation (which corresponds to $\omega\rightarrow 0$ in Eq.(\ref{G_AC_MZ_FINAL})) and find $\phi_0\equiv \arccos (\Upsilon_{DS}(0,\pi)-\Upsilon_{DS}(0,0))$, which corresponds to the phase offset at zero flux $\varphi_{AB}=0$, see Fig.(\ref{Fig3})c; (ii) we compute the phase  $\omega\vartheta= 2\arccos\left|\Upsilon_{DS}(\omega,\pi)+\Upsilon_{DS}(\omega,0)\right|$ at small $\omega$; (iii) finally, we plot Eq.(\ref{G_AC_MZ_FINAL}) using the extracted $\vartheta$ and $\psi_b(E_F)-\psi_a(E_F)=\phi_0 + \varphi_{AB}$.  We see that this formula describes quite well the numerical data, especially at low frequencies. However when the driving frequency is increased deviation of the numerical data from the fit becomes more pronounced. A more detailed analysis shows that these deviations can be accounted for by including the quadratic term in energy, which we have neglected in Eq.(\ref{Phase_F_of_E}). In Fig.~\ref{Fig3}d we plot the extracted $\omega\vartheta$ as a function of driving frequency. Various plots correspond to the samples with a different width of the hole in the interferometer. Varying this width, we modify the length of the upper path (dashed line in Fig.~\ref{Fig3}a). Extracting the corresponding slopes, i.e. $\vartheta$, for each sample, we are able to calculate the velocity of the edge state $v_g$ owing to Eq.(\ref{Tau_Theta}). For the parameters chosen in our calculation, $B\simeq 20 {\rm T}$ and the lattice constant $a_0\simeq 1 {\rm nm}$, we have obtained the velocity $v_g\simeq 0.7a_0t\hbar^{-1}$ (which corresponds to $v_g\simeq 10^6 {\rm m/s}$), where $t=\hbar^2/(2ma^2_0)$ is a hopping parameter of the tight-binding model \cite{Datta1997} used to simulate the setup, $m$\,--\,effective mass. 

In Figs.~\ref{Fig3}e,f we present the AC conductance calculations as a function of frequency. On Fig.~\ref{Fig3}e, using the source-to-drain conductance $\Upsilon_{DS}(\omega,\varphi_{AB})$, we plot the function
\begin{equation}
\label{Xi}
\Xi(\omega)=\left|\frac{\Upsilon_{DS}(\omega,\pi)-\Upsilon_{DS}(\omega,0)}{\Upsilon_{DS}(0,\pi)-\Upsilon_{DS}(0,0)}\right|,
\end{equation}
which for a simple case of Eq.(\ref{G_AC_MZ_FINAL}) reduces to $\sin(\omega\vartheta/2)/(\omega\vartheta/2)$. Again, the symbols correspond to the Green's function calculations, while the lines are given by the analytical fit, Eq.(\ref{G_AC_MZ_FINAL}), using the calculated before values of $\omega\vartheta$ (see Fig.~\ref{Fig3}d). Black circles and red diamonds represent calculations with various values of the hole width (allowing to change the length difference between the paths). We notice that for the frequencies $\hbar\omega\lesssim 0.02$, the numerical data is very well fit by Eq.(\ref{G_AC_MZ_FINAL}), while at higher values of $\hbar\omega$ this is no longer true. There are two reasons for this. First, at high enough frequencies detuning of the QPCs from the half transparency value becomes significant (see Fig.~\ref{Fig3}b) and we cannot neglect the energy dependence of the transmission/reflection amplitudes. Second, due to the dispersion of the edge state, there is always a contribution from the quadratic term neglected in the expansion (\ref{Phase_F_of_E}), which becomes important at high frequencies.

Finally, we plot the frequency dependence of the phase of the AC conductance (see Fig.~\ref{Fig3}f) varying the hole width in order to extract the second time scale, $\tau$ according to Eq.(\ref{G_AC_MZ_FINAL}). We find again the value of the group velocity $v_g\simeq 0.72a_0t\hbar^{-1}$, which is consistent with the previous result.

To conclude, we have analyzed  the AC response of the MZ electronic interferometer and found that the non-equilibrium dynamics makes it possible to address the internal time scales of the setup via transport measurements.

\section{Time dependent NEGF formalism in a nut shell}\label{Ch_TD_NEGF}
The rest of this manuscript is devoted to the derivation of the expressions given in Section \ref{Cookbook} and Appendix \ref{Cookbook_general} as well as systematic tools to derive other observables. The formalism developed in the manuscript is based on the Non Equilibrium Green's Function formalism (NEGF). While the application of NEGF to stationary quantum transport is now standard, we quickly review below the main features of its time-dependent version\cite{Jauho_Wingreen_Meir,Wingreen_Jauho_Meir,Guo_Current_partition,Guo_NonAdiab_charge_pump_Multiphoton_expansion}.
Our starting point is a general time-dependent quadratic Hamiltonian for our system:
\be
\label{Full_Hamiltonian}
{\hat{\mathbf{H}}}(t)=
\sum_{n,m}{\mathbf{H}}_{nm}(t)c^{\dagger}_nc_m.
\ee
We do not include electron-electron interactions besides some mean field treatment as was discussed in Sec. \ref{Screening}.

The basic objects that will be manipulated in this paper are various sorts of Green's Functions (GFs).
The retarded $\mathfrak{G}^{r}_{nm}(t,t')$, advanced $\mathfrak{G}^{a}_{nm}(t,t')$, lesser $\mathfrak{G}^{<}_{nm}(t,t')$ and greater $\mathfrak{G}^{>}_{nm}(t,t')$ Green's functions are defined as,
\begin{align}
&\mathfrak{G}^{r}_{nm}(t,t')=-\frac{i}{\hbar}\theta(t-t')\langle\{c_n(t),c^{\dagger}_m(t')\}\rangle\,,\label{Retarded_GF_definition}\\
&\mathfrak{G}^{a}_{nm}(t,t')=\frac{i}{\hbar}\theta(t'-t)\langle\{c_n(t),c^{\dagger}_m(t')\}\rangle\,,\label{Advanced_GF_definition}\\
&\mathfrak{G}^{<}_{nm}(t,t')=\frac{i}{\hbar}\langle c^{\dagger}_m(t')c_n(t)\rangle\,,\label{Lesser_GF_definition}\\
&\mathfrak{G}^{>}_{nm}(t,t')=-\frac{i}{\hbar}\langle c_n(t)c^{\dagger}_m(t')\rangle. \label{Greater_GF_definition}
\end{align}
where $c^{\dagger}_n(t)$ ($c_n(t)$) corresponds to $c^{\dagger}_n$ ($c_n$) in the Heisenberg representation.

The retarded and lesser/greater GFs satisfy the following equations of motion \cite{Rammer2007,Jauho2010},
\begin{align}
&\left(i\hbar\frac{\partial}{\partial t}-{\mathbf{H}}(t)\right)\mathfrak{G}^{r}(t,t')=\delta(t-t')\,,\label{Retarded_GF_EqMotion}\\
&\left(i\hbar\frac{\partial}{\partial t}-{\mathbf{H}}(t)\right)\mathfrak{G}^{\kappa}(t,t')=0,\;\kappa=<,>\,. \label{Lesser_GF_EqMotion}
\end{align}

Table \ref{TableGFs} summarizes the various Green functions introduced so far, as well as the one needed later for the AC formalism.
\begin{table}[t]
\caption{Summary of notations.}
\label{TableGFs}
\begin{ruledtabular}
\begin{tabular}{cl}
Type of GF\footnote{$\kappa=r,a,<,>$} & \multicolumn{1}{c}{Description}\\\hline
$\mathfrak{g}^{\kappa}(t,t')$ & GF of the system when the leads and the\\
& scattering region are decoupled, see Eq.(\ref{Hamiltonian_splitting}).\\\hline
$g^{\kappa}(t,t')$ & GF of the device sub-block for a system with\\ &decoupled leads. $g^{\kappa}(t,t')\equiv \mathfrak{g}_{\bar 0 \bar 0}^{\kappa}(t,t')$\\\hline\hline
$\mathfrak{G}^{\kappa}(t,t')$ & GF of the system described by the full\\
& Hamiltonian (\ref{Full_Hamiltonian}).\\\hline
$G^{\kappa}(t,t')$ & GF of the device sub-block, see Eq.(\ref{Device_Region_Notation}).\\
&$G^{\kappa}(t,t')\equiv\mathfrak{G}_{\bar 0\bar 0}^{\kappa}(t,t')$\\\hline
$G^{\kappa}_l(E)$ & GF of the device sub-block with $l$ photons\\
& emitted/absorbed, see Eq.(\ref{GF_Fourier_trans}).\\\hline
$G^{(n)}_l(E)$ & n-th order in $V_{ac}$ of the device sub-block GF\\
&with $l$ photons emitted/absorbed, see Eq.(\ref{GF_expansion}).\\\hline
$\mathcal{G}_l(E)$ & Retarded \textit{equilibrium} GF of the system at\\
&energy $E+\frac{\hbar\omega l}{2}$, see Eqs.(\ref{G_0_definition}) and (\ref{Eq_GF_notation}).\\
\end{tabular}
\end{ruledtabular}
\end{table}

\subsection{Dyson equation}

It is often convenient to split the full Hamiltonian (\ref{Full_Hamiltonian}) into a sum of an
"unperturbed part" $\mathcal{H}$ and a perturbation $\mathcal{V}(t)$:
\begin{equation}
{\mathbf{H}}=\mathcal{H}+\mathcal{V}(t).\label{Hamiltonian_splitting}
\end{equation}
This splitting can (and will) be done in several different ways, dictated by convenience. For instance, $\mathcal{V}$ can be a time-dependent potential, or a hopping element between the device and the leads, or a sum of the previous two, etc. 
Introducing $\mathfrak{g}^{r}(t,t')$ and $\mathfrak{g}^{<}(t,t')$, the unperturbed Green's functions associated to $\mathcal{H}$, one can derive
 the Dyson equations \cite{Wingreen_Meir,Rammer_Smith,Rammer2007,Jauho2010}, which relate the full GFs to the unperturbed ones. They read,
 \begin{align}
\mathfrak{G}^r&=\mathfrak{g}^r+\mathfrak{g}^r\ast \mathcal{V}\ast \mathfrak{G}^r\,,\label{Dyson_equation_1}\\
\mathfrak{G}^r&=\mathfrak{g}^r+\mathfrak{G}^r\ast \mathcal{V}\ast \mathfrak{g}^r, \label{Dyson_equation_2}\\
\mathfrak{G}^{\kappa}&=\mathfrak{g}^{\kappa}+\mathfrak{g}^{r}\ast \mathcal{V}\ast \mathfrak{G}^{\kappa}+\mathfrak{g}^{\kappa}\ast \mathcal{V}\ast \mathfrak{G}^{a},\;\kappa=<,>\,,\label{Lesser_GF_Keldysh_1}\\
\mathfrak{G}^{\kappa}&=\mathfrak{g}^{\kappa}+\mathfrak{G}^{r}\ast \mathcal{V}\ast \mathfrak{g}^{\kappa}+\mathfrak{G}^{\kappa}\ast \mathcal{V}\ast \mathfrak{g}^{a},\;\kappa=<,>\,,\label{Lesser_GF_Keldysh_2}
\end{align}
where the symbol $\ast$ stands for convolution with respect to time and matrix product with respect to the site indices:
\be
(A\ast B)_{ij}(t,t')=\sum_k \int dt^{''}A_{ik}(t,t^{''})B_{kj}(t^{''},t')
\ee
and $\mathcal{V}$ should be understood as $\delta(t-t')\mathcal{V}(t)$ in a convolution.

\subsection{Integrating out the electrodes}\label{G_lesser_section}
From now on we restrict $\mathcal{V}(t)$ to the matrix elements that couple the leads to the system plus (possibly) a time dependent potential in the device region. Then from Eqs.(\ref{Dyson_equation_1}) and (\ref{Lesser_GF_Keldysh_1}) one arrives at

\begin{equation}
G^{r}=g^{r}+g^{r}\ast(\Sigma^r+V)\ast G^{r},\label{Retarded_GF_eqn_integrated_leads}
\end{equation}
and
\begin{align}
G^{\kappa}&=g^{\kappa}+g^{r}\ast
V\ast G^{\kappa}+g^{r}\ast
\Sigma^{r}\ast G^{\kappa}+g^{r}\ast \Sigma^{\kappa}\ast G^{a}\notag\\
&+g^{\kappa}\ast V\ast G^{a}
+g^{\kappa}\ast\Sigma^{a}\ast G^{a},\;\kappa=<,>\,,\label{Lesser_GF_eqn_integrated_leads}
\end{align}
where we have introduced special notations for the $(\bar{0}\bar{0})$ device sub-block (see Section \ref{DC_primer}),
\begin{equation}
H\equiv\mathcal{H}_{\bar{0}\bar{0}},\;
V=\mathcal{V}_{\bar{0}\bar{0}},\;
G^{\kappa}\equiv \mathfrak{G}^{\kappa}_{\bar{0}\bar{0}},\;
g^{\kappa}\equiv \mathfrak{g}^{\kappa}_{\bar{0}\bar{0}},\;
...
\label{Device_Region_Notation}
\end{equation}
and also the self-energies $\Sigma^{\kappa}$ defined as
\begin{align}
\begin{split}
\Sigma^{\kappa}&(m)=\mathcal{V}_{\bar{0}\bar{m}}\ast \mathfrak{g}^{\kappa}_{\bar{m}\bar{m}}\ast \mathcal{V}_{\bar{m}\bar{0}},\\
\Sigma^{\kappa}&=\sum_{m=1}^{M}\Sigma^{\kappa}(m),\;\kappa=r,a,<,>.\label{Retarded_SE_definition}
\end{split}
\end{align}
Utilizing Eq.(\ref{Retarded_GF_EqMotion}), equation (\ref{Retarded_GF_eqn_integrated_leads}) can be rewritten in terms of an effective equation of motion,
\begin{equation}
\left(i\hbar\frac{\partial}{\partial t}-H\right)G^{r}(t,t')-(\Sigma^{r}+V)\ast G^{r}=\delta(t-t'),\label{Retarded_GF_EqMotion_integrated}
\end{equation}
while the lesser and greater GFs (\ref{Lesser_GF_eqn_integrated_leads}) with the help of Eqs.(\ref{Retarded_GF_EqMotion})-(\ref{Lesser_GF_EqMotion}) and (\ref{Retarded_GF_EqMotion_integrated}) are simplified to,
\begin{equation}
G^{\kappa}=G^{r}\ast\Sigma^{\kappa}\ast G^{a},\;\kappa=<,>.\label{Lesser_GF_NoLeads}
\end{equation}

In the absence of  time-dependent perturbations in the Hamiltonian (\ref{Hamiltonian_splitting}), the Fourrier transform of the 
self-energies $\Sigma^{<}(m)$ and $\Sigma^{>}(m)$ are given by fluctuation-dissipation theorem  which results in\cite{Datta1997,Rammer2007,Jauho2010},
\begin{align}
\mathsf{\Sigma}^{<}(m;E)&=if_m(E)\Gamma_m(E),\label{SelfEn_lesser_equilibrium}\\
\mathsf{\Sigma}^{>}(m;E)&=-i(1-f_m(E))\Gamma_m(E),\label{SelfEn_greater_equilibrium}
\end{align}
with $\Gamma_m$ defined in Eq.(\ref{Gamma_def}).

\subsection{Expression for the current}

The current associated with the $m$-th lead is found using the approach of \cite{Wingreen_Meir,Jauho_Wingreen_Meir,Wingreen_Jauho_Meir,Rammer_Smith}, i.e. calculating the change of the number of particles in the lead due to connection with the device region and thereby with other leads too. So, the expression for the current reads
\begin{equation}
I_m=-e\langle\frac{d\hat{N}_m}{dt}\rangle,\;\hat{N}_m=\sum_{\alpha\in\sigma_m}c^{\dagger}_{\alpha}(t)c_{\alpha}(t).\label{Current_definition}
\end{equation}
Taking into account the definition (\ref{Lesser_GF_definition}), Eq.(\ref{Current_definition}) transforms into
\begin{align}
I_m(t)=e\!\!\sum_{\substack{i\in\sigma_0,\\ \alpha\in\sigma_{m}}}\left(\mathcal{V}_{\alpha i}(t)\mathfrak{G}^{<}_{i \alpha}(t,t)-\mathcal{V}_{i \alpha}(t)\mathfrak{G}^{<}_{\alpha i}(t,t)\right).\label{Current_equation}
\end{align}
Let us introduce auxiliary quantity,
\begin{align}
J(m;t,t')=&e\!\!\sum_{\substack{i\in\sigma_0,\\ \alpha\in\sigma_{m}}}\left(\mathcal{V}_{\alpha i}(t')\mathfrak{G}^{<}_{i \alpha}(t,t')-\mathcal{V}_{i \alpha}(t)\mathfrak{G}^{<}_{\alpha i}(t,t')\right)\notag \\
\equiv& e\textrm{Tr}\left(\mathfrak{G}^{<}_{\bar{0}\bar{m}}\ast \mathcal{V}_{\bar{m}\bar{0}}-
\mathcal{V}_{\bar{0}\bar{m}}\ast \mathfrak{G}^{<}_{\bar{m}\bar{0}}\right).\label{Current_equation_1}
\end{align}
Exploiting Eqs.(\ref{Lesser_GF_Keldysh_1}) and (\ref{Lesser_GF_Keldysh_2}), one can get
\begin{align}
\mathfrak{G}^{<}_{\bar{m}\bar{0}}=\mathfrak{g}^{r}_{\bar{m}\bar{m}}\ast \mathcal{V}_{\bar{m}\bar{0}}\ast \mathfrak{G}^{<}_{\bar{0}\bar{0}}+\mathfrak{g}^{<}_{\bar{m}\bar{m}}\ast \mathcal{V}_{\bar{m}\bar{0}}\ast \mathfrak{G}^{a}_{\bar{0}\bar{0}},\label{Lesser_GF_LeadDev}\\
\mathfrak{G}^{<}_{\bar{0}\bar{m}}=\mathfrak{G}^{r}_{\bar{0}\bar{0}}\ast \mathcal{V}_{\bar{0}\bar{m}}\ast \mathfrak{g}^{<}_{\bar{m}\bar{m}}+\mathfrak{G}^{<}_{\bar{0}\bar{0}}\ast \mathcal{V}_{\bar{0}\bar{m}}\ast \mathfrak{g}^{a}_{\bar{m}\bar{m}}.\label{Lesser_GF_DevLead}
\end{align}
Substituting expressions (\ref{Lesser_GF_LeadDev}) and (\ref{Lesser_GF_DevLead}) into Eq.(\ref{Current_equation_1}), we come to
\begin{align}
\begin{split}
J(m;t,t')=e\textrm{Tr}\left(G^{r}\ast\Sigma^{<}(m)+
G^{<}\ast\Sigma^{a}(m)\right.\\
\left.-\Sigma^{r}(m)\ast G^{<}-
\Sigma^{<}(m)\ast G^{a}\right).\label{Current_general}
\end{split}
\end{align}
Having this, we can easily find our expression for the current, since
\begin{equation}
I_m(t)=J(m;t,t).
\end{equation}

\section{Effect of a periodic potential}\label{Ch_Periodic_potential}

We now use explicitly the fact that the perturbation (\ref{W}) is periodic in time. We introduce the Wigner coordinates $\{\tau=t-t',T=(t+t')/2\}$ and notice that the GF is periodic in $T$, $G^{r}(\tau,T)\rightarrow G^{r}(\tau,T+2\pi/\omega)$. Thus it is possible to expand the GF into a Fourier series with respect to $T$ and into a Fourier integral with respect to $\tau$ \cite{Guo_NonAdiab_charge_pump_Multiphoton_expansion,Aguado_Platero,Tsuji_PRB2008}
\begin{equation}
G^{r}(\tau,T)=\int\frac{dE}{2\pi\hbar}\sum_{l=-\infty}^{\infty}e^{-\frac{i}{\hbar}E\tau}e^{-i\omega lT}G^r_l(E). \label{GF_Fourier_trans}
\end{equation}
We will use extensivley the fact that when $C(t,t')=A\ast B$ is  a convolution of two quantities, one gets \cite{Tsuji_PRB2008},
\begin{align}
C_l(E)=\sum_{l_1+l_2=l}A_{l_1}(E+\frac{\hbar\omega l_2}{2})&B_{l_2}(E-\frac{\hbar\omega l_1}{2}).\label{ConvolTrans}
\end{align}
For instance, the currents reads,
\begin{equation}
I_m(t)=\int\frac{dE}{2\pi\hbar}\sum_{l=-\infty}^{\infty}e^{-i\omega lt}J_l(m;E),\label{Current_FourierTransformation}
\end{equation}
where using Eq.(\ref{ConvolTrans}) for each term in Eq.(\ref{Current_general}) we arrive at
\begin{widetext}
\begin{align}
J_l(m;E)=e\!\!\sum_{l_1+l_2=l}\!\!\textrm{Tr}\left[G^r_{l_1}(E+\frac{\hbar\omega l_2}{2})\Sigma^{<}_{l_2}(m;E-\frac{\hbar\omega l_1}{2})+
G^{<}_{l_1}(E+\frac{\hbar\omega l_2}{2})\Sigma^{a}_{l_2}(m;E-\frac{\hbar\omega l_1}{2})\right.\notag\\
\left.-\Sigma^{r}_{l_1}(m;E+\frac{\hbar\omega l_2}{2})G^{<}_{l_2}(E-\frac{\hbar\omega l_1}{2})-
\Sigma^{<}_{l_1}(m;E+\frac{\hbar\omega l_2}{2})G^{a}_{l_2}(E-\frac{\hbar\omega l_1}{2})\right].\label{Current_FourierImage}
\end{align}
\end{widetext}
This is the main starting point of all the subsequent derivations. We still have to supplement it by three equations. First, using Eqs.(\ref{Lesser_GF_NoLeads}) and (\ref{ConvolTrans}) it is straightforward to find
\begin{align}
G^{<}_l(E)=\sum_{m'=1}^{M}\sum_{l_1+l_2+l_3=l}G^{r}_{l_1}(E+\frac{\hbar\omega l_2}{2}+\frac{\hbar\omega l_3}{2})\times\notag\\
\Sigma^{<}_{l_2}(m';E-\frac{\hbar\omega l_1}{2}+\frac{\hbar\omega l_3}{2})G^{a}_{l_3}(E-\frac{\hbar\omega l_1}{2}-\frac{\hbar\omega l_2}{2}).
\end{align}
Second, if we consider the definition of the advanced and the retarded GFs and a symmetry of the transformation (\ref{GF_Fourier_trans}), we easily get
\begin{align}
G^{a}_l(E)=&\left[G^{r}_{-l}(E)\right]^{\dagger}.
\end{align}
Finally, with the two previous expressions and Eq.(\ref{Current_FourierImage}) one can deduce
\begin{align}
J_l(m;E)=[J_{-l}(m;E)]^{\dagger}.
\end{align}

In the next sections we develop a systematic way to calculate the current (\ref{Current_FourierTransformation}), i.e. to express
the GF elements $G^{r}_{l}(E)$ in terms of known quantities. We consider two type of perturbations: internal perturbation (in the device region) and external perturbation (in the contacts). 

\section{Perturbation in the device region}\label{Ch_Pert_device}
In this section we consider the case when the perturbation is applied inside the scattering region. Therefore we assume that the leads are at local thermal equilibrium and unaffected by the perturbation. This implies that $V$ in Eq.(\ref{Retarded_GF_EqMotion_integrated}) is given by Eqs.(\ref{W}) and (\ref{W_device}). Performing the transformation (\ref{GF_Fourier_trans}) in the Eq.(\ref{Retarded_GF_EqMotion_integrated}) we get
\begin{align}
&\left(E+\frac{\hbar\omega l}{2}-H-\mathsf{\Sigma}^r(E+\frac{\hbar\omega l}{2})\right)G^r_l(E)\notag\\
&-eV_{ac}\frac{W}{2}G^r_{l-1}(E-\frac{\hbar\omega}{2})-
eV_{ac}\frac{W}{2}G^r_{l+1}(E+\frac{\hbar\omega}{2})=\delta_{l,0}\label{GF_EqMotion_AC_E}
\end{align}
or equivalently,
\begin{align}
&G^r_l(E)-eV_{ac}{\mathcal{G}}_l(E)\frac{W}{2}\left(G^r_{l-1}(E-\frac{\hbar\omega}{2})\right.\notag\\
&\left. +G^r_{l+1}(E+\frac{\hbar\omega}{2})\right)=\delta_{l,0}{\mathcal{G}}_0(E).\label{GF_EqMotion_AC_E_modified}
\end{align}
In the two following paragraphs we explore two complementary limits: small perturbation amplitude ($eV_{ac}$) and adiabatic limit ($\omega\rightarrow 0$). 
\subsection{Limit of a small perturbation amplitude}
\begin{figure}[h]
\begin{center}
\includegraphics[scale=0.8]{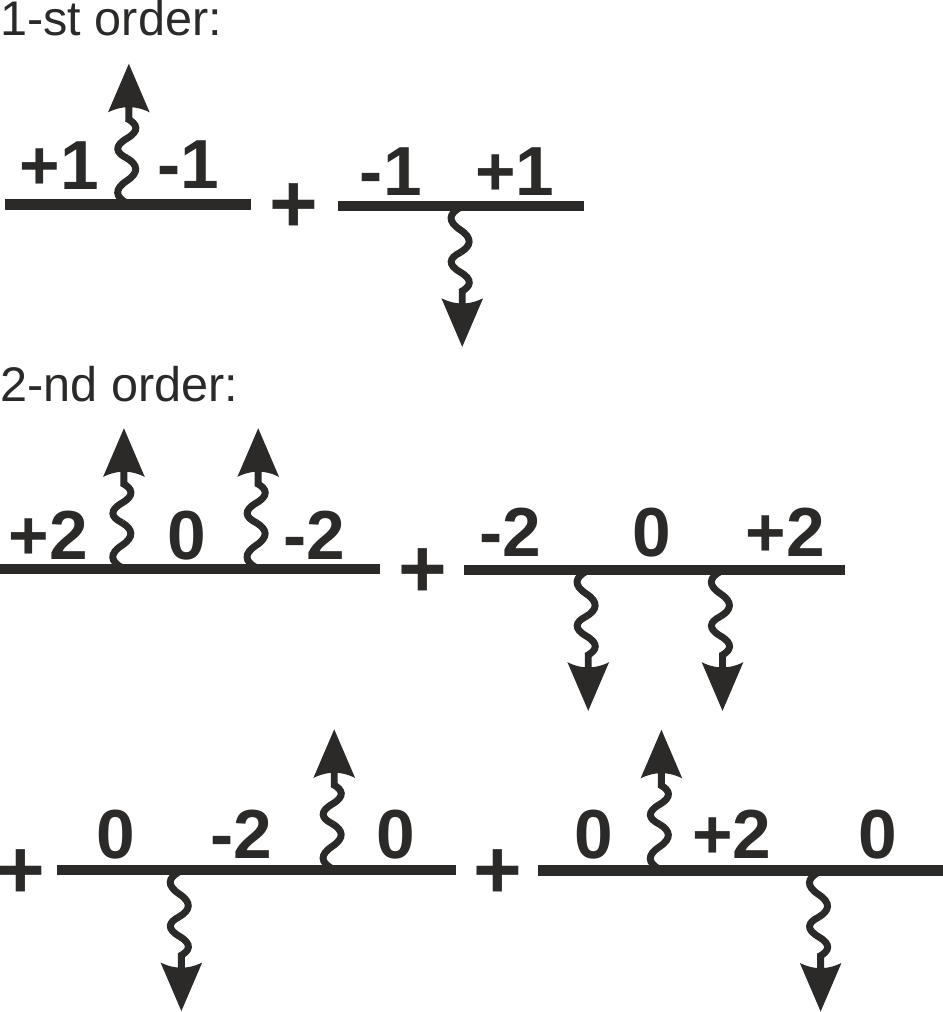}
\caption{\label{Diagrams_1_2_orders} First and second order diagrams. Numbers over the straight lines correspond to the index $l$ for $\mathcal{G}_l$. Direction of the wavy line tells us whether the absorbtion or emission of a photon takes place. Each diagram corresponds directly to one term in Eqs.(\ref{AC_System_first_order},\ref{AC_System_second_order})}
\end{center}
\end{figure}

Let us assume that the perturbation amplitude is much smaller than all characteristic energy scales in the system, e.g. hopping constant between sites. Then we can solve Eq.(\ref{GF_EqMotion_AC_E}) iteratively in powers of $eV_{ac}$. The solution takes form
\begin{align}
G^r_l(E)=\sum_{n=0}^{\infty}(eV_{ac})^nG^{(n)}_l(E)&,\label{GF_expansion}\\
G^{(n)}_l(E)={\mathcal{G}}_l(E)\frac{W}{2}\left(G^{(n-1)}_{l-1}(E-\frac{\hbar\omega}{2})\right.&\left.+G^{(n-1)}_{l+1}(E+\frac{\hbar\omega}{2})\right),
\end{align}
for $n\geq1$ and
\begin{align}
G^{(0)}_l(E)=\delta_{l,0}{\mathcal{G}}_l(E).\label{AC_System_zero_order}
\end{align}
This equation can be solved iteratively. It is instructive to write down explicitly the first and second order contributions,
\begin{align}
G^{(1)}_l(E)=&\delta_{l,1}{\mathcal{G}}_1(E)\frac{W}{2}{\mathcal{G}}_{-1}(E)+\delta_{l,-1}{\mathcal{G}}_{-1}(E)\frac{W}{2}{\mathcal{G}}_{1}(E),\label{AC_System_first_order}\\
G^{(2)}_l(E)=&\delta_{l,2}{\mathcal{G}}_{2}(E)\frac{W}{2}{\mathcal{G}}_{0}(E)\frac{W}{2}{\mathcal{G}}_{-2}(E)\notag\\
+&\delta_{l,-2}{\mathcal{G}}_{-2}(E)\frac{W}{2}{\mathcal{G}}_{0}(E)\frac{W}{2}{\mathcal{G}}_{2}(E)\notag\\
+&\delta_{l,0}\left({\mathcal{G}}_{0}(E)\frac{W}{2}{\mathcal{G}}_{-2}(E)\frac{W}{2}{\mathcal{G}}_{0}(E)\right.\notag\\
+&\left.{\mathcal{G}}_{0}(E)\frac{W}{2}{\mathcal{G}}_{2}(E)\frac{W}{2}{\mathcal{G}}_{0}(E)\right)\label{AC_System_second_order}
\end{align}
The structure of the two previous equations suggests the following simple diagrammatic representation of an arbitrary order contribution
(see Fig.~\ref{Diagrams_1_2_orders}). The diagrams are made by horizontal propagating lines $[{\mathcal{G}}_{l}(E)]$ separated by 
"photon absorption/mission" vertical wavy lines $[W/2]$. In order to build $G^{(n)}_l(E)$ one has to remember the following "Feynman rules",
\begin{itemize} 
\item To get the contributions of order $n$, draw $n$ wavy lines pointing up or down in all possible configurations (there are $2^n$ diagrams).
\item Each diagram of order $n$ gives a contribution to $G^{(n)}_l(E)$ where $l$ is the difference between the number of up wavy lines
and down wavy lines.
\item Read the diagram from left to right. Starting from ${\mathcal{G}}_{l}(E)$, each wavy line correspond to a factor $W/2$ followed
by another ${\mathcal{G}}_{l'}(E)$ with $l'$ decreased by 2 (up wavy line, a "photon" is emitted) or increased by 2 (down wavy line, "photon" absorbed). Repeat until the end of the diagram.
\end{itemize}
\subsection{Adiabatic limit and beyond}\label{Adiabatic_device}
We now turn to a very different limit where the driving frequency (and not the amplitude) is the small parameter of the problem. When the perturbation varies slowly the system follows adiabatically. We introduce the generating function $F$ as,
\begin{equation}
F(z,E)\equiv \sum_{l=-\infty}^{\infty}z^{l} G_l(E+\frac{\hbar\omega l}{2}).\label{Generating_function}
\end{equation}
A closed equation for $F$ can be obtained by expanding the self energy,
\begin{align}
&\mathsf{\Sigma}^{r}(E+\hbar\omega l)= \mathsf{\Sigma}^{r}(E)+\sum_{n=1}^\infty\frac{1}{n!}(\hbar\omega l)^n
\frac{\partial^n \mathsf{\Sigma}^{r}}{\partial E^n}.
\label{WBL-Adiabatic}
\end{align}
Assuming that we work around the wide band limit and using Eq.(\ref{GF_EqMotion_AC_E}) (evaluated at energy $E+\hbar\omega l/2$), we obtain up to first order in the derivative of $\mathsf{\Sigma}^{r}(E)$,
\begin{equation}
F + \hbar\omega z F_{ad} \left[ 1 - \frac{\partial \mathsf{\Sigma}^{r}}{\partial E} \right] \frac{\partial F}{\partial z}  = F_{ad}\label{eqF}
\end{equation}
with 
\begin{equation}
F_{ad}=\frac{1}{E-H-\Sigma^r(E)-eV_{ac}\frac{W}{2}\left(z+\frac{1}{z}\right)}.\label{Fad}
\end{equation}
Note that $F_{ad}$ corresponds to the adiabatic limit: when one evaluates F for a given time $T$ (see Eq.(\ref{GF_Fourier_trans})), $F_{ad}(e^{-i\omega T},E)$ corresponds to the stationary retarded Green's function
at energy $E$ for the potential at time $T$, i.e. assuming that at a given time $T$, the potential varies so slowly that it can be considered as constant. Higher order terms can be obtained straightforwardly and
correspond to higher derivatives of $F$. For instance, to second order, one should add the following to the left hand side of 
Eq.(\ref{eqF})
\begin{equation}
-\frac{(\hbar\omega)^2}{2}   F_{ad}(z) \frac{\partial^2 \mathsf{\Sigma}^{r}}{\partial E^2}  \left[z^2 \frac{\partial^2 F}{\partial z^2} + z \frac{\partial F}{\partial z}\right].
\end{equation}
Equation (\ref{eqF}) allows for a systematic calculation of $F$ (and therefore the $G_l$), for instance by expanding it in powers of $\hbar\omega$. To first order we get,
\begin{equation}
F(e^{-i\omega T})=F_{ad}(e^{-i\omega T})-i\hbar F_{ad}(e^{-i\omega T}) \left[ 1 - \frac{\partial \mathsf{\Sigma}^{r}}{\partial E} \right] \frac{\partial F_{ad}}{\partial T}.
 \end{equation}
And higher orders are obtained straightforwardly. We emphasize that in the adiabatic limit the processes contain a (arbitrary) large number of absorbed/emitted photons, hence the role of the generating functions which correspond to an instantaneous basis. The resulting observables are given in the cookbook section \ref{Cookbook} and in  Appendix~\ref{Cookbook_general}.

\section{Perturbation in the leads}\label{Ch_Pert_leads}
The formalism developed above can be extended to homogeneous perturbations in the leads. The algebra is very similar with one
notable exception: multiple absorption/emission processes are now allowed. We suppose (for definitness) that a bias voltage $V_{ac} \cos\omega t$ is applied to lead $\bar m'$, see Eq.(\ref{W_leads}).

\subsection{Equation of motion}
It is convenient to change the basis in the lead affected by the perturbation in order to move to a frame where the lead is stationary.
The AC voltage then gives rise to a time-dependent phase factor in the coupling matrix between the lead and the device.
 This is easily seen with the help of the unitary transformation
\begin{equation}
\hat{U}=\exp\left[{\displaystyle\frac{i}{\hbar}\int_0^{t}dt'
eV_{ac}\cos(\omega t')\hat{N}}\right], \hat{N}=\sum_{i\in\sigma_m}c^{\dagger}_{\alpha}c_{\alpha}.\label{Gauge_transformation}
\end{equation}
The Hamiltonian after the transformation refers to the old one,
$$
\mathbf{H}'=\hat{U}\,\mathbf{H}\hat{U}^{\dagger}-i\hbar\,\hat{U}\frac{\partial \hat{U}^{\dagger}}{\partial t}.
$$
Then, it consists of
\begin{equation}
\mathbf{H}'(t)=\mathcal{H}+\mathcal{V}(t),
\end{equation}
where $\mathcal{H}$ is the Hamiltonian of the leads and device (when they are decoupled) and,
\begin{align}
\mathcal{V}(t)=\sum_{m=1}^{M}\sum_{\alpha\in\sigma_m,i\in\sigma_0}
(&\mathcal{V}_{\alpha i}e^{\frac{ieV_{ac}}{\hbar\omega}\sin\omega t}c^{\dagger}_{\alpha}c_i\notag\\
+&\mathcal{V}_{i \alpha}e^{-\frac{ieV_{ac}}{\hbar\omega}\sin\omega t}c^{\dagger}_ic_{{\alpha}})
\end{align}
is the \textit{new} coupling between them. By doing this change of basis we are back to the situation where the leads are kept at (local) thermal equilibrium, whereas the effect of the perturbation is completely transferred to the coupling matrix between the latter and the scattering region. As a result, the self energies of lead $\bar m'$ now acquire an additional phase factor,
\begin{equation}
\Sigma^{\kappa}(m';t,t')=\mathsf{\Sigma}^{\kappa}(m';t-t')e^{-\frac{ieV_{ac}}{\hbar\omega}(\sin\omega t-\sin\omega t')},
\;\kappa=r,a,<,\label{Self_Energy_LeadsPert}
\end{equation}
where $\mathsf{\Sigma}^{\kappa}(m';t-t')$ is the equilibrium self-energy (in the absence of the AC field).
Expanding the phase factor in Eq.(\ref{Self_Energy_LeadsPert}) in terms of Bessel functions,
\begin{align}
e^{-\frac{ieV_{ac}}{\hbar\omega}\sin\omega t}\notag =\sum_{n=-\infty}^{+\infty} J_n \left(\frac{eV_{ac}}{\hbar\omega}\right) 
e^{-i\omega nt},
\end{align}
the transformation (\ref{GF_Fourier_trans}) applied to the (perturbed) self-energy (\ref{Self_Energy_LeadsPert}) gives
\begin{align}
&\Sigma^{\kappa}_l(m';E)\notag\\
&=\!\!\sum_{n=-\infty}^{\infty}\!\!\!J_{l+n}\!\!\left(\frac{eV_{ac}}{\hbar\omega}\right)\!\!
J_n\!\!\left(\frac{eV_{ac}}{\hbar\omega}\right)\!\!\mathsf{\Sigma}^{\kappa}\!\!\left(m';\!E-\frac{\hbar\omega l}{2}-\hbar\omega n\!\right)\!\!.\label{SelfEn_perturbated_lead}
\end{align}
Finally, the equation of motion has the form
\begin{align}
&\left(E+\frac{\hbar\omega l}{2}-H-\sum_{m\neq m'}^M\mathsf{\Sigma}^{r}(m;E+\frac{\hbar\omega l}{2})\right)G^r_l(E)\notag\\
&-\!\!\sum_{l_1=-\infty}^{\infty}\Sigma^{r}_{l_1}\!\!\left(m';E+\frac{\hbar\omega}{2}(l-l_1)\right)\!G^r_{l-l_1}\!\left(E-\frac{\hbar\omega l_1}{2}\right)\!\!=\delta_{l,0}.\label{EqMotion_Gret_in_leads}
\end{align}
This equation is the starting point for the approximation schemes considered below.
\subsection{Limit of a small perturbation amplitude}

Let us now expand $G^r_l(E)$ in powers of $eV_{ac}/\hbar\omega\ll 1$,
\begin{align}
G^r_l(E)=\sum_{n=0}^{\infty}\left(\frac{eV_{ac}}{\hbar\omega}\right)^nG^{(n)}_l(E).
\end{align}
The first order result can be obtained by a direct expansion of Eq.(\ref{SelfEn_perturbated_lead}),
\begin{align}
&\Sigma^{\kappa}_l(m';E)=\mathsf{\Sigma}^{\kappa}(m';E)\delta_{l,0}\notag\\
&+\frac{eV_{ac}}{2\hbar\omega}\left[\mathsf{\Sigma}^{\kappa}(m';E-\frac{\hbar\omega}{2})-\mathsf{\Sigma}^{\kappa}(m';E+\frac{\hbar\omega}{2})\right]\delta_{l,1}\notag\\
&+\frac{eV_{ac}}{2\hbar\omega}\left[\mathsf{\Sigma}^{\kappa}(m';E-\frac{\hbar\omega}{2})-\mathsf{\Sigma}^{\kappa}(m';E+\frac{\hbar\omega}{2})\right]\delta_{l,-1}\notag\\
&+O\left(e^2V^2_{ac}\right),\;\kappa=r,a,<,
\end{align}
where we used the power series representation of Bessel functions. Utilizing the notation introduced in Eq.(\ref{Delta_x_x}), 
we obtain
\begin{align}
&G^{(0)}_l(E)=\left(\frac{1}{2}\right)^0\delta_{l,0}\mathcal{G}_0,\\
\begin{split}
G^{(1)}_l(E)=\left(\frac{1}{2}\right)^1\left[\delta_{l,-1}\mathcal{G}_{-1}\Lambda^{rr}_{m'}(E-\frac{\hbar\omega}{2};E+\frac{\hbar\omega}{2})\mathcal{G}_{1}\right.\\
+\left.\delta_{l,1}\mathcal{G}_{1}\Lambda^{rr}_{m'}(E-\frac{\hbar\omega}{2};E+\frac{\hbar\omega}{2})\mathcal{G}_{-1}\right].\label{GF_1_leads}
\end{split}
\end{align}
which is similar to expressions (\ref{AC_System_zero_order}) and (\ref{AC_System_first_order}).

In analogy with the case of internal perturbations, a systematic diagrammatic expansion can be constructed. The set of rules
to obtain all the contributions to $G^{(n)}_l(E)$ is given by,  
\begin{itemize} 
\item Draw all diagrams with $1\leq p \leq n$ wavy lines. Contrary to the previous case, the wavy lines now point both up {\it and} down, reflecting the possibility of multiple absorption and emission processes. Each wavy line $i$ is associated with two positive integers $n_i^a$ and $n_i^e$  ($n_i^e+n_i^a\ge 1$) that correspond to the two types of processes. We have
\begin{align}
n=&\sum_{i=1}^p (n_i^e+n_i^a),\\
l=&\sum_{i=1}^p (n_i^e-n_i^a).
\end{align} 
\item Read the diagram from left to right. Starting from ${\mathcal{G}}_{l}(E)$, each wavy line corresponds to a factor $T^l_{n_i^a,n_i^e}$
(see below) followed by another ${\mathcal{G}}_{l'}(E)$ with $l'=l+2(n_i^a - n_i^e)$  ($n_i^e$ "photons" are emitted and $n_i^a$ are absorbed). Repeat until the end of the diagram. 
\item The "vertex" is given by
\begin{equation}
\label{T_element}
\begin{split}
T_{n_i^a,n_i^e}^l=\frac{1}{2^{n_i^e +n_i^a}}\frac{1}{n_i^e!n_i^a!}\\
\times\Delta^{n_i^e +n_i^a}_{m'}\left(E+\frac{\hbar\omega}{2}[l+n_i^a -n_i^e]\right),
\end{split}
\end{equation}
with the matrix $\Delta^{q}_{m'}(E)$ given by
\begin{equation}
\begin{split}
\Delta^{q}_{m'}(E)=\sum_{i=0}^{q-1} (-1)^i \combinatorial{q-1}{i} \\
\times\Lambda^{rr}_{m'}\left(E+(2i-q)\frac{\hbar\omega}{2};E+(2i+2-q)\frac{\hbar\omega}{2}\right),
\end{split}
\end{equation}
\end{itemize}
where $\combinatorial{q}{i}$ is the binomial coefficient.
\begin{figure}[]
\begin{center}
\includegraphics[angle=0,width=0.7\linewidth]{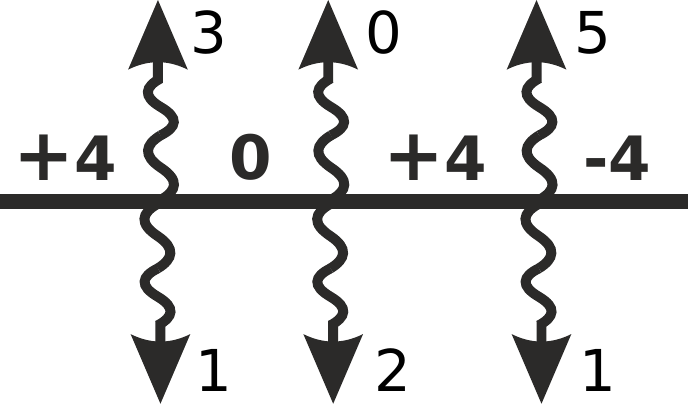}
\caption{Example of a diagram that contributes to $G^{(12)}_4(E)$. Each diagram is characterized by the set of upper and lower numbers which correspond respectively to the number of emitted and absorbed photons. The numbers along the horizontal line are calculated from the diagrammatic rules.}
\label{Diagram_leads_example}
\end{center}
\end{figure}
As an example, Fig.~\ref{Diagram_leads_example} corresponds to one contribution to the $12$-th order with $l=4$, namely $G^{(12)}_4(E)$. Using the above rules, this diagram gives,
\begin{align}
G^{(12)}_4(E)=\mathcal{G}_{4}T^4_{1,3}\mathcal{G}_{0}T^0_{2,0}\mathcal{G}_{4}T^4_{1,5}\mathcal{G}_{-4},
\end{align}
where, according to Eq.(\ref{T_element}),
\begin{align}
&T^4_{1,3}=\frac{1}{2^43!}\left[\Lambda^{rr}_{m'}\left(E-\hbar\omega ;E\right)-3\Lambda^{rr}_{m'}\left(E;E+\hbar\omega\right)\right.\notag\\
&\left. +3\Lambda^{rr}_{m'}\left(E+\hbar\omega ;E+2\hbar\omega\right)
-\Lambda^{rr}_{m'}\left(E+2\hbar\omega ;E+3\hbar\omega\right)\right],\\
&T^0_{2,0}=\frac{1}{2^22!}\left[\Lambda^{rr}_{m'}\left(E;E+\hbar\omega\right)-\Lambda^{rr}_{m'}\left(E+\hbar\omega ;E+2\hbar\omega\right)\right],\\
&T^4_{1,5}=\frac{1}{2^65!}\left[\Lambda^{rr}_{m'}\left(E-3\hbar\omega ;E-2\hbar\omega\right)\right.\notag\\
&-5\Lambda^{rr}_{m'}\left(E-2\hbar\omega;E-\hbar\omega\right)+10\Lambda^{rr}_{m'}\left(E-\hbar\omega ;E\right)\notag\\
&-10\Lambda^{rr}_{m'}\left(E;E+\hbar\omega\right)+5\Lambda^{rr}_{m'}\left(E+\hbar\omega; E+2\hbar\omega\right)\notag\\
&\left.-\Lambda^{rr}_{m'}\left(E+2\hbar\omega; E+3\hbar\omega\right)\right].
\end{align}

 We see that in contrast to the case when the perturbation was inside the scattering region, multiple-photon absorption/emission processes are allowed. This fact can be also understood from the concept of the \textit{sidebands} (with energy shifted with respect to the Fermi level by an amount $\pm n\hbar\omega$) which have been introduced in the context of AC scattering theory\cite{Tien_Gordon,Trauzettel_Blanter,Aguado_Platero}.

\subsection{Adiabatic limit and beyond}\label{Adiabatic_leads}
Finally, we consider the adiabatic limit following a similar procedure to the one presented in section \ref{Adiabatic_device}.
 The procedure is very similar except for the expansion of the self energy of the lead under AC perturbation. 
 The general expansion of Eq.(\ref{SelfEn_perturbated_lead}) reads,
\begin{widetext}
\begin{align}
\Sigma^{r}_{l_1}\!\!\left(m';E+\frac{\hbar\omega}{2}(l-l_1)\right)=
\sum_{k=0}^{\infty}\frac{1}{k!}\left(\frac{\hbar\omega}{2}\right)^{k}\frac{\partial^{k}\mathsf{\Sigma}^{r}(m';E)}{\partial E^{k}}\sum_{n=-\infty}^{\infty}\!\!\!(l-2l_1-2n)^{k}J_{l_1+n}\left(\frac{eV_{ac}}{\hbar\omega}\right)J_n\left(\frac{eV_{ac}}{\hbar\omega}\right). \label{SelfEn_perturbated_lead_adiabatic}
\end{align}
\end{widetext}
which, using the following two identities for Bessel functions,
\begin{align}
&\sum_{n=-\infty}^{\infty}J_{n+l}(x)J_{n}(x)=\delta_{l,0},\\
&\frac{2n}{x}J_n(x)=J_{n-1}(x)+J_{n-1}(x),
\end{align}
allows to obtain the self energy to any value of $k$. Restricting to first order, we get
\begin{align}
&\Sigma^{r}_{l_1}\!\!\left(m';E+\frac{\hbar\omega}{2}(l-l_1)\right)\approx\delta_{l_1,0}\mathsf{\Sigma}^{r}(m';E)\notag\\
&-\frac{eV_{ac}}{2}(\delta_{l_1,1}+\delta_{l_1,-1})\frac{\partial\mathsf{\Sigma}^{r}(m';E)}{\partial E}+\delta_{l_1,0}\frac{\hbar\omega l}{2}\frac{\partial\mathsf{\Sigma}^{r}(m';E)}{\partial E}.\notag
\end{align}
This expansion corresponds to the wide-band limit ($k=0$)\cite{Jauho_Wingreen_Meir,Guo_AC_conductance_NTs} and its first correction ($k=1$). It is expected to be very accurate in metallic leads, for instance. At this level, we introduce again the generating function for
$F(z)$, Eq.(\ref{Generating_function}) and obtain the same equation \ref{eqF} as for the internal perturbation case provided one replace $W$ by $[-{\partial\mathsf{\Sigma}^{r}(m';E)}/ {\partial E}]$. The corresponding results can hence be adapted to this case straightforwardly.
Note that beyond this first order closed equation can also be obtained, but this simple replacement rule does not apply anymore.

\section{Conclusions}\label{Ch_Conclusions}

Numerical simulations of quantum transport has become an ubiquitous tool for mesoscopic physics and are more and more
commonly used to help the design of nanoelectronic devices. On the other hand there is a general trend of mesoscopic physics and
microelectronics towards GHz or even higher frequencies, so that developing a general framework to tackle finite frequency transport
is becoming of increasing importance. In this manuscript we have developed the corresponding formalism allowing to derive a large set
of formulas that express AC observables in terms of numerically accessible quantities. We provide systematic rules to construct other
expressions that we did not give explicitly. Hence, this manuscript can either be used as a recipe book for extending DC numerical tools to AC, or as a starting point for further developments.
\bibliography{References}
\bibliographystyle{apsrev}

\appendix

\section{General results for AC observables given in Section \ref{Cookbook}}\label{Cookbook_general}
In this Appendix we present the formulas for AC observables beyond the WBL, see Section \ref{Cookbook}. Just as before we will split them in two parts: response to the internal and external perturbations.
\subsection{External perturbation}
First, we consider the situation when the AC voltage is applied to one of the contacts, say $m'$. Below we present various system response functions to such a perturbation assuming there is no DC bias in the system.

The \textit{AC conductance matrix} is given by
\begin{widetext}
\begin{align}
\label{AC_conductance_outside_general}
\begin{split}
\Upsilon_{m,m'}(\omega)=\frac{e^2}{h}\int\frac{dE}{\hbar\omega}\Biggl\{ \left(f(E)-f(E+\hbar\omega)\right)\mathrm{Tr}\left[\Lambda^{ar}_m(E;E+\hbar\omega)\mathcal{G}_{2}\Lambda^{ra}_{m'}(E+\hbar\omega;E)\mathcal{G}^{\dagger}_{0}\right]\\
-f(E)\mathrm{Tr}\left[\Lambda^{rr}_m(E;E+\hbar\omega)\mathcal{G}_{2}\Lambda^{rr}_{m'}(E+\hbar\omega;E)\mathcal{G}_{0}\right]+f(E+\hbar\omega)\mathrm{Tr}\left[\Lambda^{aa}_m(E;E+\hbar\omega)\mathcal{G}^{\dagger}_{2}\Lambda^{aa}_{m'}(E+\hbar\omega;E)\mathcal{G}^{\dagger}_{0}\right]\\
-\mathrm{Tr}\left[f(E)D_1-f(E+\hbar\omega)D_2\right]\delta_{m,m'}\Biggr\}.
\end{split}
\end{align}
\end{widetext} 
Functions $D_{1,2}$ in the diagonal part are defined as
\begin{align}
D_1=i\left(\mathcal{G}_{2}-\mathcal{G}^{\dagger}_{0}\right)\Gamma(m';E)+\left(\mathcal{G}_{0}-\mathcal{G}^{\dagger}_{0}\right)\Lambda^{rr}_{m'}(E;E+\hbar\omega),\label{D1}
\end{align}
\begin{align}
\begin{split}
D_2=i\left(\mathcal{G}_{2}-\mathcal{G}^{\dagger}_{0}\right)\Gamma(m';E+\hbar\omega)\\
+\left(\mathcal{G}_{2}-\mathcal{G}^{\dagger}_{2}\right)\Lambda^{aa}_{m'}(E;E+\hbar\omega).\label{D2}
\end{split}
\end{align}

The \textit{stationary (DC) component of the current} to leading order in perturbation amplitude reads ($m\neq m'$)
\begin{widetext}
\begin{align}
\label{rectification}
\begin{split}
\frac{d^2I_m(0\omega)}{dV^2_{ac}}=\frac{ie^3}{4\hbar^2\omega^2}\int \frac{dE}{2\pi\hbar}\Bigl\{\left(f(E)-f(E+\hbar\omega)\right)\mathrm{Tr}\left[\mathcal{G}_{0}\Lambda^{ra}_{m'}(E;E+\hbar\omega)\mathcal{G}^{\dagger}_{2}\Lambda^{aa}_{m'}(E+\hbar\omega;E)\mathcal{G}^{\dagger}_{0}\Gamma_m(E)\right.\\
\left.-\mathcal{G}_{0}\Lambda^{rr}_{m'}(E;E+\hbar\omega)\mathcal{G}_{2}\Lambda^{ra}_{m'}(E+\hbar\omega;E)\mathcal{G}^{\dagger}_{0}\Gamma_m(E)\right]\\
+\left(f(E-\hbar\omega)-f(E)\right)\mathrm{Tr}\left[
\mathcal{G}_{0}\Lambda^{rr}_{m'}(E;E-\hbar\omega)\mathcal{G}_{-2}\Lambda^{ra}_{m'}(E-\hbar\omega;E)\mathcal{G}^{\dagger}_{0}\Gamma_m(E)\right.\\
\left.-\mathcal{G}_{0}\Lambda^{ra}_{m'}(E;E-\hbar\omega)\mathcal{G}^{\dagger}_{-2}\Lambda^{aa}_{m'}(E-\hbar\omega;E)\mathcal{G}^{\dagger}_{0}\Gamma_m(E)\right]\\
-if(E)\mathrm{Tr}\left[\mathcal{G}_{0}\Bigl(\Gamma_{m'}(E+\hbar\omega)+\Gamma_{m'}(E-\hbar\omega)-2\Gamma_{m'}(E)\Bigr)\mathcal{G}^{\dagger}_0\Gamma_m(E)\right]\Bigr\}.
\end{split}
\end{align}
\end{widetext}

\textit{Adiabatic current.} Assuming that the frequency of the perturbation is very small, to zeroth order in $\hbar\omega$ the current is ($m\neq m'$)
\begin{widetext}
\begin{align}
\label{adiabatic}
\begin{split}
I^{ad}_m(t)=\frac{e}{h}\int dE\Bigl\{f(E)\mathrm{Tr}\left[i\left(F^{(0)}(t,E)-F^{(0),\dagger}(t,E)\right)\Gamma_m(E)-F^{(0)}(t,E)\Gamma(E)F^{(0),\dagger}(t,E)\Gamma_m(E)\right]\\
+\mathrm{Tr}\left[F^{(0)}(t,E)\left(\frac{\partial\Gamma_{m'}(E)}{\partial E}f(E)+\frac{\partial f}{\partial E}\Gamma_{m'}(E)\right)F^{(0),\dagger}(t,E)\Gamma_m(E)\right]eV_{ac}\cos\omega t\Bigr\},
\end{split}
\end{align}
\end{widetext}
where the definition of the functions $F^{(0),(1)}(t,E)$ are given by,
\begin{equation}
F^{(0)}(t,E)=\frac{1}{E-H-\Sigma^r(E)+\frac{\partial\mathsf{\Sigma}^{r}(m';E)}{\partial E}eV_{ac}\cos\omega t},\label{F0_Pert_leads}
\end{equation}
\begin{align}
\begin{split}
F^{(1)}(t,E)=i F^{(0)}\left[\frac{\partial\mathsf{\Sigma}^{r}(m')}{\partial E}F^{(0)}
\left(1-\frac{\partial \mathsf{\Sigma}^{r}}{\partial E}\right)\right.\\
\left.-\left(1-\frac{\partial \mathsf{\Sigma}^{r}}{\partial E}\right)F^{(0)}
\frac{\partial\mathsf{\Sigma}^{r}(m')}{\partial E}\right]F^{(0)}\frac{eV_{ac}}{2}\sin\omega t.\label{F1_Pert_leads}
\end{split}
\end{align}
\textit{Correction to the adiabatic current.} It is the leading order correction of the order $\sim\hbar\omega$ to Eq.(\ref{adiabatic}). Note that it contains all orders in $V_{ac}$,
\begin{align}
\delta I^{ad}_m(t)=\delta I^1_m(t)+\delta I^2_m(t)+\delta I^3_m(t)+\delta I^4_m(t)+\delta I^5_m(t),
\end{align}
where separate parts read,
\begin{widetext}
\begin{align}
\begin{split}
\delta I^1_m(t)=e\int\frac{dE}{2\pi\hbar}\Biggl\{f(E)\mathrm{Tr}\left[i\left(F^{(1)}(t,E)-F^{(1),\dagger}(t,E)\right)\Gamma_m(E)-F^{(1)}(t,E)\Gamma(E)F^{(0),\dagger}(t,E)\Gamma_m(E)\right.\\
-\left.F^{(0)}(t,E)\Gamma(E)F^{(1),\dagger}(t,E)\Gamma_m(E)\right]+\mathrm{Tr}\left[F^{(1)}(t,E)\left(\frac{\partial\Gamma_{m'}(E)}{\partial E}f(E)+\frac{\partial f}{\partial E}\Gamma_{m'}(E)\right)F^{(0),\dagger}(t,E)\Gamma_m(E)\right.\\
\left.+F^{(0)}(t,E)\left(\frac{\partial\Gamma_{m'}(E)}{\partial E}f(E)+\frac{\partial f}{\partial E}\Gamma_{m'}(E)\right)F^{(1),\dagger}(t,E)\Gamma_m(E)\right]\Biggr\},
\end{split}
\end{align}
\begin{align}
\begin{split}
\delta I^{2}_m(t)=\frac{e\hbar}{2}\int\frac{dE}{2\pi\hbar}\mathrm{Tr}\left[\frac{\partial(F^{(0)}(t,E)+F^{(0),\dagger}(t,E))}{\partial t}\left(\frac{\partial\Gamma_m(E)}{\partial E}f(E)+\frac{\partial f}{\partial E}\Gamma_m(E)\right)\right],
\end{split}
\end{align}
\begin{align}
\begin{split}
\delta I^{3}_m(t)=\frac{e\hbar}{2}\int\frac{dE}{2\pi\hbar}\frac{\partial}{\partial t}\Biggl\{\mathrm{Tr}\left[F^{(0)}(t,E)\Gamma(E)F^{(0),\dagger}(t,E)\frac{\partial}{\partial E}\left(\mathsf{\Sigma}^r(m;E)+\mathsf{\Sigma}^a(m;E)\right)\right]f(E)\\
-\mathrm{Tr}\left[F^{(0)}(t,E)\left(\frac{\partial\Gamma_{m'}(E)}{\partial E}f(E)+\frac{\partial f}{\partial E}\Gamma_{m'}(E)\right)F^{(0),\dagger}(t,E)\frac{\partial}{\partial E}\left(\mathsf{\Sigma}^r(m;E)+\mathsf{\Sigma}^a(m;E)\right)\right]eV_{ac}\cos\omega t\Biggr\},
\end{split}
\end{align}
\begin{align}
\begin{split}
\delta I^{4}_m(t)=\frac{ie\hbar}{2}\int\frac{dE}{2\pi\hbar}\Biggl\{\mathrm{Tr}\left[\frac{\partial F^{(0)}(t,E)}{\partial t}\Gamma(E)\frac{\partial F^{(0),\dagger}(t,E)}{\partial E}\Gamma_m(E)-\frac{\partial F^{(0)}(t,E)}{\partial E}\Gamma(E)\frac{\partial F^{(0),\dagger}(t,E)}{\partial t}\Gamma_m(E)\right]f(E)\\
+\mathrm{Tr}\left[\frac{\partial F^{(0)}(t,E)}{\partial t}\left(\frac{\partial\Gamma(E)}{\partial E}f(E)+\frac{\partial f}{\partial E}\Gamma(E)\right)F^{(0),\dagger}(t,E)\Gamma_m(E)\right.\\
\left.-F^{(0)}(t,E)\left(\frac{\partial\Gamma(E)}{\partial E}f(E)+\frac{\partial f}{\partial E}\Gamma(E)\right)\frac{\partial F^{(0),\dagger}(t,E)}{\partial t}\Gamma_m(E)\right]\Biggr\},
\end{split}
\end{align}
\begin{align}
\begin{split}
\delta I^{5}_m(t)=\frac{ie\hbar}{2}\int\frac{dE}{2\pi\hbar}\mathrm{Tr}\Biggl\{\frac{\partial F^{(0)}(t,E)}{\partial E}\left(\frac{\partial\Gamma_{m'}(E)}{\partial E}f(E)+\frac{\partial f}{\partial E}\Gamma_{m'}(E)\right)\left[\frac{\partial}{\partial t}\left(F^{(0),\dagger}(t,E)eV_{ac}\cos\omega t\right)\right]\Gamma_m(E)\\
-\left[\frac{\partial}{\partial t}\left(F^{(0)}(t,E)eV_{ac}\cos\omega t\right)\right]\left(\frac{\partial\Gamma_{m'}(E)}{\partial E}f(E)+\frac{\partial f}{\partial E}\Gamma_{m'}(E)\right)\frac{\partial F^{(0),\dagger}(t,E)}{\partial E}\Gamma_m(E)\Biggr\}.\end{split}
\end{align}
\end{widetext}

The \textit{generalized injectivity}\cite{Buttiker_JOP1993,Buttiker_Nuovo_Cimento,Wei_Wang_Curr_cons_NEGF_AC_Trans}, which is the density of particles on site $i$ injected from the lead $m'$ as a consequence of the oscillating electrochemical potential in it, is
\begin{widetext}
\begin{align}
\begin{split}
\frac{dn(i,1\omega,m')}{dV_{ac}}=\frac{ie}{h\omega}\int dE\left[\left(f(E)-f(E+\hbar\omega)\right)\mathcal{G}_2\Lambda_{m'}^{ra}(E+\hbar\omega;E)\mathcal{G}^{\dagger}_{0}-f(E)\mathcal{G}_2\Lambda_{m'}^{rr}(E+\hbar\omega;E)\mathcal{G}_{0}\right.\\
\left. +f(E+\hbar\omega)\mathcal{G}^{\dagger}_{2}\Lambda_{m'}^{aa}(E+\hbar\omega;E)\mathcal{G}^{\dagger}_{0}\right]_{ii}.
\end{split}
\end{align}
\end{widetext}

\subsection{Internal perturbation}
Now we will present the results for the response to an internal potential. Again, we assume that there is no DC bias applied to the system.

\textit{Current response linear in} $V_{ac}$ is given by
\begin{widetext}
\begin{align}
\label{AC_conductance_inside_general}
\begin{split}
\frac{dI_m(1\omega)}{dV_{ac}}=\frac{e^2}{h}\int dE\Biggl\{\left(f(E)-f(E+\hbar\omega)\right)\mathrm{Tr}\left[\Lambda^{ar}_m(E;E+\hbar\omega)\mathcal{G}_{2}W\mathcal{G}^{\dagger}_{0}\right]-f(E)\mathrm{Tr}\left[\Lambda^{rr}_m(E;E+\hbar\omega)\mathcal{G}_{2}W\mathcal{G}_{0}\right]\\
+f(E+\hbar\omega)\mathrm{Tr}\left[\Lambda^{aa}_m(E;E+\hbar\omega)\mathcal{G}^{\dagger}_{2}W\mathcal{G}^{\dagger}_{0}\right]\Biggr\}.
\end{split}
\end{align}
\end{widetext}

The \textit{DC component of the current} caused by the rectification effect can be written (to leading order in $V_{ac}$) as
\begin{widetext}
\begin{align}
\begin{split}
\frac{d^2I_m(0\omega)}{dV^2_{ac}}=\frac{e^3}{4h}\int dE\:\mathrm{Tr}\left[\left(f(E)-f(E+\hbar\omega)\right)\mathcal{G}_0W\mathcal{G}_2\Gamma(E+\hbar\omega)\mathcal{G}^{\dagger}_2W\mathcal{G}^{\dagger}_0\Gamma_m(E)\right.\\
\left.-\left(f(E-\hbar\omega)-f(E)\right)\mathcal{G}_0W\mathcal{G}_{-2}\Gamma(E-\hbar\omega)\mathcal{G}^{\dagger}_{-2}W\mathcal{G}^{\dagger}_0\Gamma_m(E)\right].
\end{split}
\end{align}
\end{widetext}

Similarly, the $2^{\rm nd}$ \textit{harmonics of the current} is given by
\begin{widetext}
\begin{align}
\begin{split}
\frac{d^2I_m(2\omega)}{dV^2_{ac}}=\frac{e^3}{2h}\int dE\mathrm{Tr}\Bigl[\left(f(E)-f(E+\hbar\omega)\right)\mathcal{G}_2W\mathcal{G}^{\dagger}_0W\mathcal{G}^{\dagger}_{-2}\Lambda^{ar}_m(E-\hbar\omega;E+\hbar\omega)\\
+\left(f(E-\hbar\omega)-f(E)\right)\mathcal{G}_2W\mathcal{G}_0W\mathcal{G}^{\dagger}_{-2}\Lambda^{ar}_m(E-\hbar\omega;E+\hbar\omega)+f(E+\hbar\omega)\mathcal{G}^{\dagger}_2W\mathcal{G}^{\dagger}_0W\mathcal{G}^{\dagger}_{-2}\Lambda^{aa}_m(E-\hbar\omega;E+\hbar\omega)\\
-f(E-\hbar\omega)\mathcal{G}_2W\mathcal{G}_0W\mathcal{G}_{-2}\Lambda^{rr}_m(E-\hbar\omega;E+\hbar\omega)\Bigr].
\end{split}
\end{align}
\end{widetext}

The response function to perturbing the onsite potential on site $i$ is the \textit{(generalized) emissivity}\cite{Buttiker_JOP1993,Buttiker_Nuovo_Cimento,Wei_Wang_Curr_cons_NEGF_AC_Trans} and it reads (note, that it differs from the original definition by B\"uttiker, where one has to multiply it by $1/(ie\omega)$)
\begin{widetext}
\begin{align}
\begin{split}
\frac{dI_m(1\omega)}{dV_{ii}}=\frac{e^2}{h}\int dE \left[\left(f(E)-f(E+\hbar\omega)\right)\mathcal{G}^{\dagger}_{0}\Lambda^{ar}_m(E;E+\hbar\omega)\mathcal{G}_{2}-f(E)\mathcal{G}_{0}\Lambda^{rr}_m(E;E+\hbar\omega)\mathcal{G}_{2}\right.\\
\left. +f(E+\hbar\omega)\mathcal{G}^{\dagger}_{0}\Lambda^{aa}_m(E;E+\hbar\omega)\mathcal{G}^{\dagger}_{2}\right]_{ii}.
\end{split}
\end{align}
\end{widetext}
\end{document}